\documentclass[sigconf]{acmart}

\AtBeginDocument{%
  }

\copyrightyear{2026}
\acmYear{2026}
\setcopyright{cc}
\setcctype{by}
\acmConference[HPDC '26]{The 35th International Symposium on High-Performance Parallel and Distributed Computing}{July 13--16, 2026}{Cleveland, OH, USA}
\acmBooktitle{The 35th International Symposium on High-Performance Parallel and Distributed Computing (HPDC '26), July 13--16, 2026, Cleveland, OH, USA}
\acmDOI{10.1145/3806645.3807574}
\acmISBN{979-8-4007-2640-8/2026/07}

\usepackage{enumitem}
\usepackage{algorithm}
\usepackage{algpseudocode}
\usepackage{soul}
\usepackage{float}
\usepackage{graphicx}
\usepackage{subcaption}
\usepackage[english]{babel}
\usepackage{makecell}
\usepackage{pbalance}
\usepackage{booktabs}
\usepackage{multirow}
\usepackage{subcaption}
\usepackage{tabularx}
\usepackage{threeparttable}
\begin{document}

\title{POLAR-PIC: A Holistic Framework for Matrixized PIC with Co-Designed Compute, Layout, and Communication}


\author{Yizhuo Rao}
\orcid{0009-0000-3572-6969}
\affiliation{%
    \institution{Sun Yat-Sen University}
    \city{Guangzhou}
    \country{China}
}
\email{raoyzh6@mail2.sysu.edu.cn}

\author{Xingjian Cui}
\affiliation{%
    \institution{Sun Yat-Sen University}
    \city{Guangzhou}
    \country{China}
}
\authornote{Contributed equally.}
\email{cuixj8@mail2.sysu.edu.cn}

\author{Shangzhi Pang}
\affiliation{%
    \institution{Sun Yat-Sen University}
    \city{Guangzhou}
    \country{China}
}
\email{pangshzh@mail2.sysu.edu.cn}

\author{Jiabin Xie}
\affiliation{%
    \institution{Sun Yat-Sen University}
    \city{Guangzhou}
    \country{China}
}
\email{xiejb6@mail2.sysu.edu.cn}

\author{Guangnan Feng}
\affiliation{%
    \institution{Sun Yat-Sen University}
    \city{Guangzhou}
    \country{China}
}
\email{fenggn7@mail.sysu.edu.cn}

\author{Jinhui Wei}
\affiliation{%
    \institution{Sun Yat-Sen University}
    \city{Guangzhou}
    \country{China}
}
\email{weijh28@mail2.sysu.edu.cn}

\author{Ziyan Zhang}
\affiliation{%
    \institution{Sun Yat-Sen University}
    \city{Guangzhou}
    \country{China}
}
\email{zhangzy273@mail2.sysu.edu.cn}

\author{Languang Gao}
\affiliation{%
    \institution{Sun Yat-Sen University}
    \city{Guangzhou}
    \country{China}
}
\email{gaolg@mail2.sysu.edu.cn}

\author{Zhenyu Wang}
\affiliation{%
    \institution{Institute of Plasma Physics, Chinese Academy of Sciences}
    \city{Hefei}
    \country{China}
}
\email{zhenyu.wang@ipp.ac.cn}

\author{Zhiguang Chen}
\affiliation{%
    \institution{Sun Yat-Sen University}
    \city{Guangzhou}
    \country{China}
}
\email{zhiguang.chen@nscc-gz.cn}
\authornote{Corresponding author.}

\author{Yutong Lu}
\affiliation{%
    \institution{Sun Yat-Sen University}
    \city{Guangzhou}
    \country{China}
}
\email{luyutong@mail.sysu.edu.cn}

\renewcommand{\shortauthors}{Yizhuo Rao et al.}

\begin{abstract}
    Particle-in-Cell (PIC) simulations are fundamental to plasma physics but often suffer from limited scalability due to particle–grid interaction bottlenecks and particle redistribution costs. Specifically, the particle–grid interaction computations have not taken full advantage of the emerging Matrix Processing Units (MPUs), the particle motion introduces irregular memory accesses, and the bulk-synchronous redistribution further destroys long-term data locality thereby limiting parallel efficiency.    
    To address these inefficiencies, we present \textbf{POLAR-PIC}, a co-designed framework for large-scale PIC simulations that (i) reformulates Field Interpolation into an MPU-friendly outer-product form, (ii) maintains a physically ordered particle layout to preserve memory contiguity, and (iii) overlaps particle communication with Deposition to hide redistribution overhead. 
    The evaluation on the pilot system of an Exascale supercomputer demonstrates that \textbf{POLAR-PIC} accelerates the entire particle-processing phase by up to \textbf{$10.9\times$} in uniform plasma and \textbf{$4.4\times$} in real-world laser-ion acceleration scenarios compared to the native WarpX reference pipeline on LX2. Ablation studies reveal that the speedups achieved by Interpolation and Deposition are \textbf{$8.0\times$} and \textbf{$13.2\times$}, respectively, and the asynchronous communication design sustains a \textbf{$99.1\%$} overlap ratio. In cross-platform comparisons,  \textbf{POLAR-PIC} achieves \textbf{$13.2\%$} of theoretical peak efficiency on the CPU-based LS system, while WarpX reaches \textbf{$9.6\%$} on NVIDIA A800 GPUs. Notably, the scalability evaluation demonstrates that \textbf{POLAR-PIC} maintains \textbf{$67.5\%$} weak scaling efficiency on over 2 million cores under high-migration dynamic workloads, highlighting the importance of holistic co-design for future matrix-centric HPC systems.    
       
\end{abstract}




\maketitle

\section{Introduction}

The Particle-in-Cell (PIC) method is fundamental to kinetic plasma physics and widely used in astrophysical and laboratory plasma research~\cite{bird2022vpic,fedeli2022warpx,fonseca2002osiris,arber2015epoch}. Large-scale PIC simulations, which track billions of particles while solving Maxwell's equations on Eulerian grids, impose severe pressure on the compute, memory, and interconnect subsystems. On the other hand, as heterogeneous HPC systems evolve toward the Exascale era, efficient hardware utilization has become a critical challenge for both the HPC and computational physics communities.

\begin{figure}[t]
    \centering
    \includegraphics[width=\linewidth]{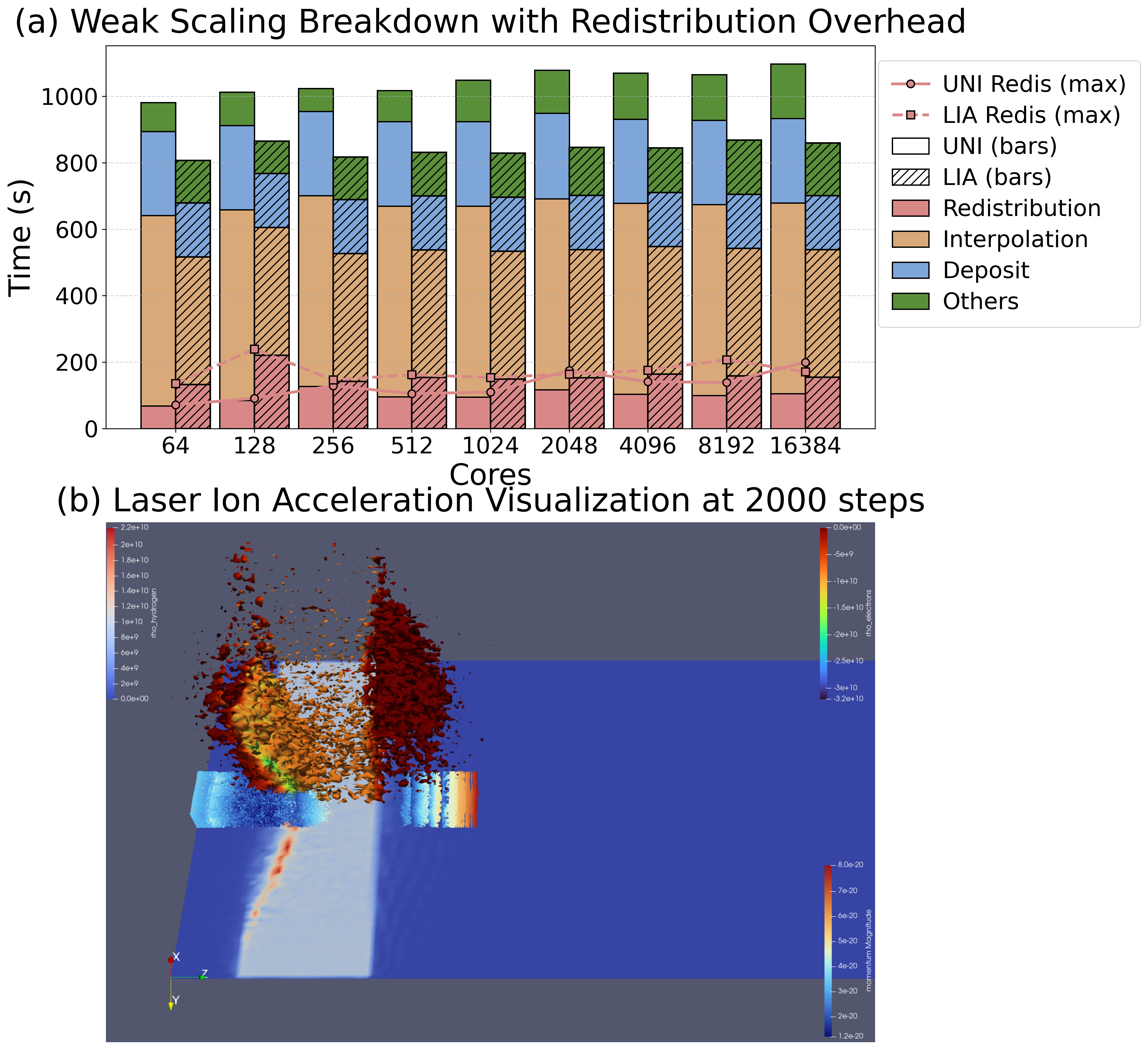} 
    \caption{Native WarpX v24.07 Runtime breakdown of Uniform Plasma and Laser-Ion Acceleration simulation on the Tianhe Xingyi platform.}
    \label{fig:pic-breakdown}
\end{figure}

Although frameworks such as VPIC~\cite{bird2022vpic} and WarpX~\cite{fedeli2022warpx, myers2021warpxgpu} scale well on heterogeneous systems, particle-grid interaction remains the dominant bottleneck, often accounting for over 80\% of runtime~\cite{bird2022vpic, fedeli2022warpx, myers2021warpxgpu, rao2026matrix} (Figure~\ref{fig:pic-breakdown}). It consists of Field Interpolation, a bandwidth-bound Gather-Stencil kernel, and Charge/Current Deposition, a Scatter-Add kernel limited by write conflicts and atomic overheads~\cite{fedeli2022warpx,yoshikawa2021fugaku}. Existing optimizations such as SoA and periodic sorting help~\cite{fedeli2022warpx, myers2021warpxgpu, rao2026matrix,derouillat2018smilei}, but irregular particle motion still prevents sustained bandwidth on VPUs and MPUs.

Specifically, modern processor architectures are rapidly evolving toward matrix-centric compute paradigms. Representative CPUs, incorporating architectural extensions such as Intel AMX~\cite{endo2026amx} and Arm SME~\cite{remke2025sme}, now integrate specialized Matrix Processing Units (MPUs). These units execute Matrix Outer-Product Accumulate (MOPA) instructions to compute and accumulate the outer product of two vectors into on-chip tiles within a single cycle, offering a quadratic increase in arithmetic density compared to VPUs. Matrix-PIC~\cite{rao2026matrix} capitalizes on this opportunity by reformulating the Deposition kernel into an outer-product form. By combining incremental particle sorting with tile-level buffering to mitigate write conflicts, it achieves significant acceleration for high-order Deposition. 

However, localized optimization of the Deposition phase is insufficient to resolve particle-processing performance bottlenecks in MPU-based PIC simulations, due to three fundamental limitations.
First, governed by Amdahl's Law, accelerating Deposition shifts the primary computational bottleneck to the Field Interpolation stage. Mathematically, Field Interpolation constitutes an inner-product reduction (many-to-one), presenting a structural mismatch with the native outer-product mechanism of MPUs (one-to-many).
Second, Matrix-PIC\cite{rao2026matrix} maintains only the logical order of particle indices, relying on intermittent global reordering to restore physical memory contiguity. As particle disorder accumulates, the growing reordering cost introduces significant runtime overhead and performance jitter.
Third, mainstream frameworks such as WarpX/AMReX~\cite{fedeli2022warpx,zhang2019amrex} typically employ bulk-synchronous parallel (BSP) execution models, treating particle redistribution as a global synchronization point. While these frameworks exhibit robust weak scalability, the BSP pattern constrains single-step performance, particularly in high-dynamic, non-uniform scenarios like Laser-Ion Acceleration (LIA). As shown in Figure~\ref{fig:pic-breakdown}, the redistribution cost in LIA escalates with increasing parallelism, identifying particle redistribution as an expanding bottleneck at scale.

To address the challenges of operator structural mismatch, fragmented data supply, and blocking synchronization, we propose \textbf{POLAR-PIC} (\textbf{P}article \textbf{O}uter-product, \textbf{L}ocality-\textbf{A}ware, and \textbf{R}MA-overlapped \textbf{PIC}). This framework co-designs kernel reformulation, data layout maintenance, and communication scheduling to elevate the optimization scope from isolated kernel acceleration to particle-processing critical path reduction on MPU-based architectures. The primary contributions of this paper are organized as follows:

\begin{enumerate}[label=\arabic*., leftmargin=10pt, labelsep=0.5em, itemsep=2pt]
  \item \textbf{We introduce a novel co-design of a matrix outer-product formulation for the PIC Field Interpolation operator.} By transforming the Field Interpolation operator into an MPU-compatible outer-product form via a cell-centric batching strategy, we bridge the structural mismatch between the Interpolation's native inner-product logic and the hardware's outer-product primitives, mitigating the bottleneck shift predicted by Amdahl's Law.
  
  \item \textbf{We propose an inline Sort-on-Write (SoW) mechanism to prevent physical layout degradation arising from the divergence between logical indexing and physical storage.} Rather than performing a periodic global sort, SoW leverages the inherent read-modify-write path and thread-local tile buffering to maintain strict cell-grouped physical contiguity for the majority of particles with negligible amortized $O(1)$ overhead, ensuring a stable, contiguous data supply for matrix-based operations.

  \item \textbf{We design a fine-grained communication overlap strategy using pre-packing and notifiable RMA.} To mitigate synchronization stalls caused by particle redistribution, we fuse the packing of migrating particles directly into the Field Interpolation kernel and initiate non-blocking transfers using the high-performance one-sided communication library UNR~\cite{feng2024unr}. This scheduling effectively hides packing overhead, overlaps particle communication with the Deposition kernel, and optimizes the entire particle communication path.
\end{enumerate}

In summary, \textbf{POLAR-PIC} synergizes operator outer-product reformulation, the SoW locality maintenance mechanism, and RMA-overlapped communication into a cohesive optimization paradigm that advances PIC simulations toward next-generation HPC systems. Built upon the open-source, highly optimized WarpX\cite{fedeli2022warpx}, our experiments on the next-generation HPC platform equipped with an LX2 CPU demonstrate that \textbf{POLAR-PIC} accelerates the particle-processing phase by up to $10.9\times$ in uniform plasma and $4.4 \times$ in laser-ion acceleration scenarios over the WarpX baseline. Furthermore, it surpasses the matrix-based SOTA Matrix-PIC by $4.7\times$ and $3.8\times$ in these respective scenarios. Moreover, we report a cross-platform peak-efficiency comparison for context, where \textbf{POLAR-PIC} attains $13.2\%$ of theoretical peak efficiency on the LS pilot system and WarpX reaches $9.6\%$ on NVIDIA A800.


\section{Related Work} 

\subsection{Vectorization and Matrixization in PIC} 

As processor architectures evolve toward heterogeneous designs, exploiting SIMD techniques is increasingly necessary for maximizing PIC compute throughput. Early architectural optimizations focused on vector processing units (VPUs), leveraging data-layout transformations (AoS-to-SoA)~\cite{decyk2014emerging}, particle binning~\cite{decyk2014emerging}, and portable VPU backends in production codes~\cite{bird2022vpic,fedeli2022warpx,myers2021warpxgpu,zhang2019amrex}, with adaptive SIMD strategies to accommodate nonuniform particle distributions~\cite{derouillat2018smilei}. Nevertheless, irregular Gather/Scatter behavior remains largely constrained by bandwidth and address-generation overheads, a limitation also observed when mapping sparse/irregular kernels to accelerators~\cite{hong2018spmm}, while similar refactor-and-vectorize practices have benefited other large-scale scientific codes~\cite{gao2022Optimization}. 
The subsequent emergence of matrix-centric hardware has driven a paradigm shift toward mapping structured-grid stencils onto high-density Matrix Outer-Product Accumulate (MOPA) primitives~\cite{huang2025stencil,zhang2024LoRAStencil,chen2024ConvStencil,liu2022stencil_gpu}, alongside layout-aware redesigns in scientific models~\cite{cao2023AGCM-3DLF} and even matrix-centric formulations of compression algorithms~\cite{song2024CereSZ}. In particle simulations, Matrix-PIC~\cite{rao2026matrix} pioneered refactoring the Deposition kernel into an MPU-friendly outer-product formulation on emerging matrix-centric CPUs, opening a CPU-oriented path that is complementary to mature GPU-based PIC pipelines. Rather than targeting a direct replacement of existing GPU solutions, this line of work explores how matrix-oriented computation, locality maintenance, and runtime scheduling can be co-designed within a unified general-purpose execution environment. However, Matrix-PIC leaves Field Interpolation unresolved due to the structural mismatch between Interpolation’s inner-product reduction and the native outer-product mechanism of MPUs. Consequently, effectively mapping massive particle Gather to MPUs remains an open gap on next-generation matrix architectures; removing this emerging Amdahl bottleneck is the efficiency gap that this work aims to close.
\vspace{-8pt}
\subsection{Data Layout and Particle Sorting}

PIC performance hinges on maintaining spatio-temporal locality against particle diffusion~\cite{unat2017trends,beck2019AdaptiveSIMD}, typically necessitating SoA layouts~\cite{barsamian2018layouts,decyk2014emerging,beck2019AdaptiveSIMD} and Space-Filling Curves (SFC)~\cite{barsamian2018layouts}. To counter dynamic locality degradation, production frameworks employ periodic full physical sorting to restore memory compaction~\cite{myers2021warpxgpu,bird2022vpic, Barsamian2018Efficient}. However, this incurs prohibitive $O(N)$ data movement costs that preclude high-frequency execution. Alternatively, lightweight index-based strategies, such as the GPMA in Matrix-PIC~\cite{rao2026matrix,Bender2007gpma,brian2021gpma_graph} or incremental binning~\cite{derouillat2018smilei,beck2019AdaptiveSIMD}, reduce movement overheads but may not guarantee physical contiguity, leading to memory fragmentation that starves bandwidth-sensitive MPU pipelines. To resolve this tension between sorting cost and memory contiguity, POLAR-PIC adopts a Controlled Data Duplication mechanism inspired by optimization techniques in sparse matrix and graph processing~\cite{hong2018spmm,balaji2019Data_Duplication}. This approach preserves the sustained physical contiguity required for matrix acceleration while mitigating the excessive overhead of full global reordering.

\subsection{Computation-Communication Overlap}

While computation-communication overlap is a canonical strategy for scalability, standard MPI non-blocking primitives often suffer from pseudo-asynchronous behavior due to the lack of independent background progress~\cite{hoefler2008overlapping,cools2017communication-hiding,denis2022overlapMPI,zhou2024MPI_Progress,huang2023Accelerating_MPI}. To overcome these software stack limitations, Remote Direct Memory Access (RDMA) and One-sided Communication mechanisms have been developed to offload data transfer to hardware, featuring optimizations for intra-node copy engines~\cite{cho2023exploiting}, distributed locks~\cite{Schmid2016rma}, host memory management~\cite{tang2023HM2}, and event-driven notification as seen in Unified Notifiable RMA (UNR)~\cite{feng2024unr}. However, existing overlap schemes predominantly focus on static grid Halo exchange~\cite{Schaller2016SWIFT,Guidotti2021PICAsynchronous}, leaving irregular, data-dependent particle redistribution and its packing/transfer constrained by the end-of-step batching model inherent to bulk-synchronous parallel (BSP) designs. POLAR-PIC fills this gap by embedding particle packing and transmission into the computational data path, thereby avoiding explicit end-of-step synchronization on the critical path and reducing tail effects caused by migration-related overheads.

\section{Preliminaries}

This section formalizes the mathematical model of Field Interpolation and analyzes its architectural bottlenecks to motivate the proposed MPU-based optimization and SoW design.

\subsection{Field Interpolation and Particle Update}


Field Interpolation maps electromagnetic fields from discrete grid points to continuous particle positions. Mathematically, this operation represents a high-dimensional tensor contraction process, where the weight tensor derived from shape functions is contracted with the grid field tensor along the stencil dimension to resolve the field value at particle coordinates.

For a particle located at $\mathbf{x}_p = (x, y, z)$, the interpolated physical quantity $F_p$ (such as $E_p$ or $B_p$) is derived by weighting grid node values within the local support stencil. Assuming a shape function of order $S$, we have:
\begin{equation}
F(\mathbf{x}_p) = \sum_{k=0}^{S} \sum_{j=0}^{S} \sum_{i=0}^{S} \left( S_x(i)\, S_y(j)\, S_z(k) \right)\cdot 
F_{\mathbf{grid}}(i_0+i,\, j_0+j,\, k_0+k).
\label{eq:gather}
\end{equation}
where $S_x, S_y, S_z$ denote the normalized shape factors along the axes, and $(i_0, j_0, k_0)$ represents the base index of the grid cell containing the particle. The order $S$ dictates the stencil width, necessitating that the summation indices $i, j, k$ traverse all $S+1$ spatially coupled grid nodes starting from the base anchor $(i_0, j_0, k_0)$.

While Field Interpolation constitutes a memory-bound gather-stencil bottleneck on next-generation architectures due to its structural mismatch with MPU primitives, the subsequent \textbf{Particle Push} phase exposes an unavoidable write-back path that we leverage for layout maintenance. Using the interpolated fields $\mathbf{E}_p$ and $\mathbf{B}_p$, this kernel updates particle momentum and position via the Lorentz force equation~\cite{boris1970relativistic,vay2008simulation}. Crucially, the push operation follows a deterministic \textit{read-modify-write} dataflow: it streams current attributes to compute the next state and commits the results to memory. This mandatory write-back path serves as the pivotal leverage point for the layout optimizations proposed in this work, enabling dynamic reordering without additional memory passes.

\subsection{Matrix Outer-Product Calculation Paradigm}
To overcome the scaling limitations of traditional vector architectures, recent CPUs increasingly integrate Matrix Processing Units (MPUs). Distinct from Vector Processing Units (VPUs), MPUs natively support the Matrix Outer-Product Accumulate (MOPA) operation. Given $\mathbf{a} \in \mathbb{R}^m$ and $\mathbf{b} \in \mathbb{R}^n$, MOPA computes their outer product and accumulates it into an on-chip 2D tile register $\mathbf{C} \in \mathbb{R}^{m \times n}$:
\begin{equation}
\mathbf{C} \leftarrow \mathbf{C} + \mathbf{a} \otimes \mathbf{b}.
\label{eq:mopa}
\end{equation}
Representative implementations like Arm SME FMOPA deliver a near-quadratic increase in compute density by executing $m\times n$ fused multiply-add updates operations per instruction. However, the irregular access patterns inherent to PIC simulations impede pipeline saturation, necessitating paradigm migration and locality optimizations to regularize data supply for effective MPU utilization.

\subsection{PIC Workflow and Redistribution Overheads}

\begin{figure}[h]
    \centering
    \includegraphics[width=\linewidth]{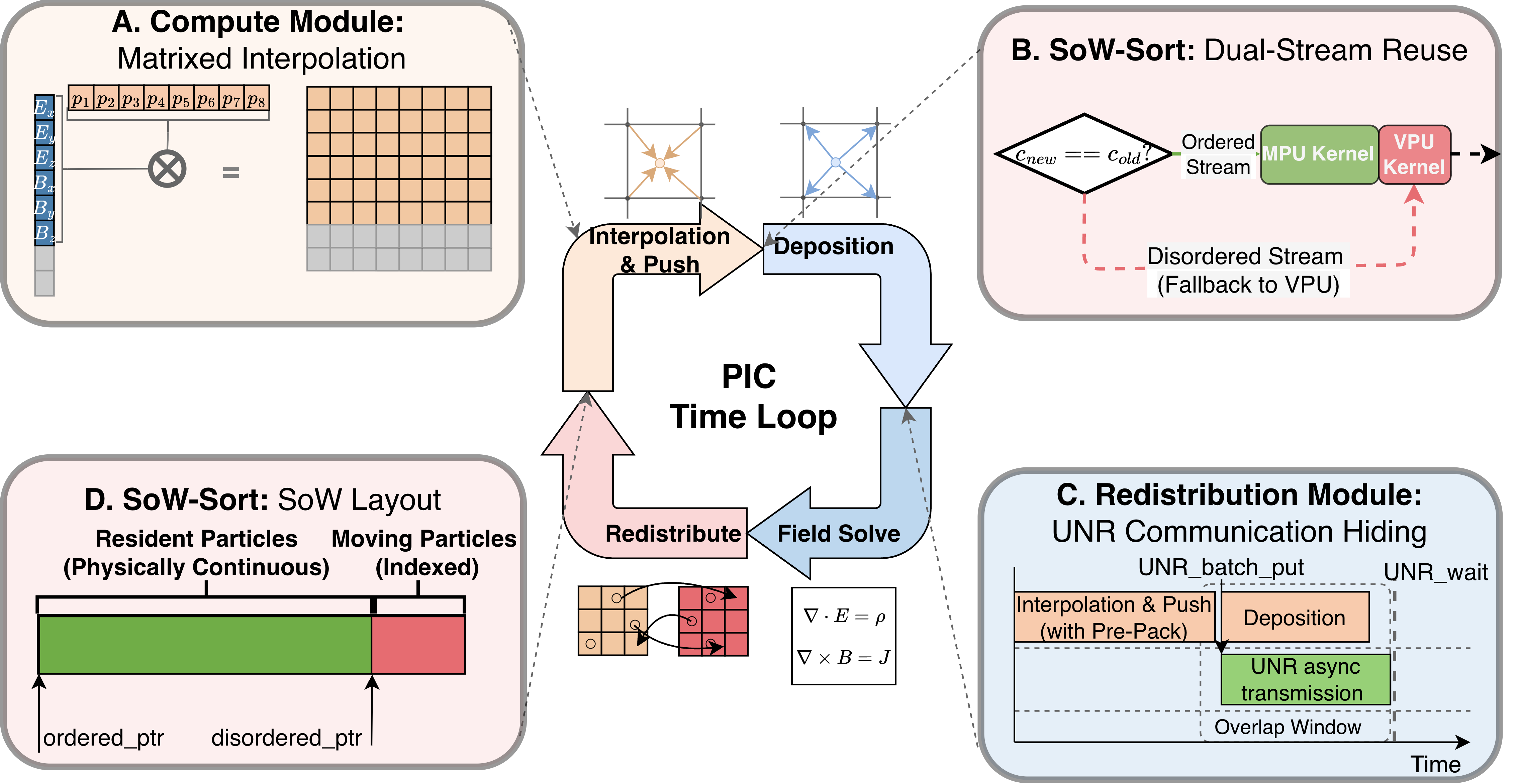}
    \caption{POLAR-PIC integrates the MPU-based Interpolation, the Dual-Stream Reuse, the UNR-based asynchronous communication pipeline, and SoW Layout.}
    \label{fig:framework}
\end{figure}

As illustrated in the central loop of Figure~\ref{fig:framework}, a standard electromagnetic PIC timestep cycles through four distinct phases: \textit{Interpolation \& Push}, \textit{Deposition}, \textit{Field Solve}, and \textit{Particle Redistribute}.

Conventionally (e.g., in WarpX~\cite{fedeli2022warpx}), the Particle Redistribute phase is positioned at the end of the timestep as a bulk-synchronous parallel (BSP) barrier. This process strictly sequences three steps: scanning and packing boundary-crossing particles, blocking communication, and unpacking. In high-dynamic scenarios, the substantial overhead of particle packing combined with this blocking communication pattern becomes a dominant performance bottleneck.
\begin{figure*}[t]
\centering
\includegraphics[width=\linewidth]{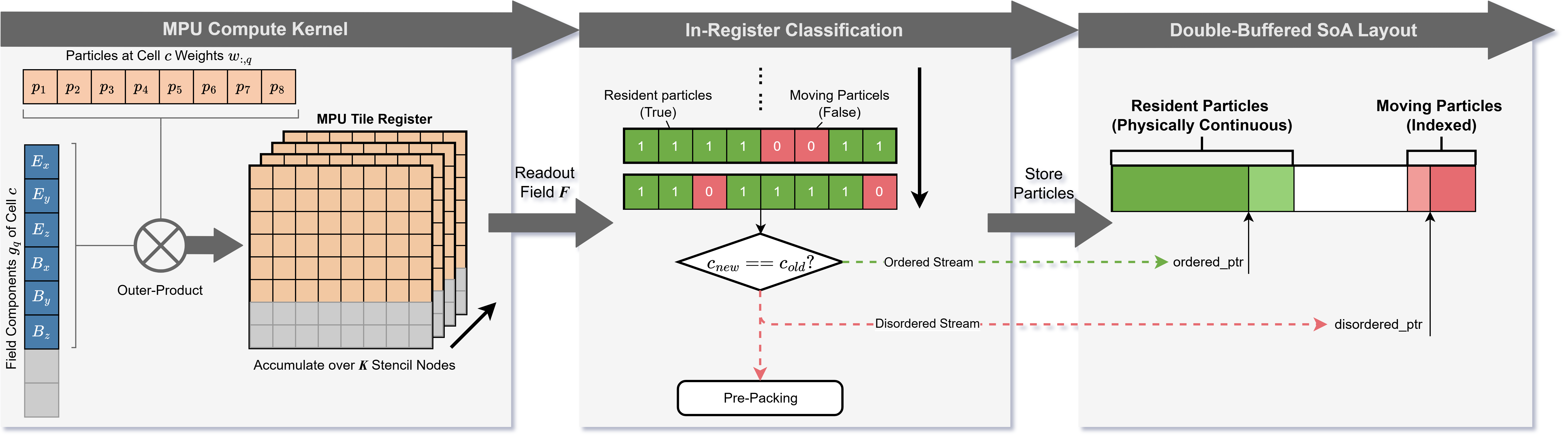} 
\caption{\textbf{Matrixized Field Interpolation and Sort-on-Write dataflow.}
Left: Matrix outer-product Interpolation via tensor stacking;
Middle: in-register classification and pre-packing;
Right: double-buffered SoA layout that maintains a physically contiguous layout.}
\label{fig:gather_tile}
\end{figure*}
\subsection{UNR Asynchronous Communication Library}

The Unified Notifiable RMA (UNR) library~\cite{feng2024unr} provides a lightweight asynchronous interface for HPC, offering one-sided RMA built on low-level communication primitives and enhanced completion notification: \texttt{UNR\_Put}/\texttt{Get} offload transfers to the NIC via RDMA with minimal CPU involvement, while notifiable completion uses native event counters to signal data arrival without explicit handshakes or matching, jointly enabling high-performance asynchronous communication suitable for tight compute pipelines in PIC simulations.

\section{Methodology}

This section details the system design of POLAR-PIC. Guided by the characteristics of next-generation architectures, we propose a co-optimization scheme that integrates an outer-product Field Interpolation operator, a Sort-on-Write (SoW) layout maintenance strategy, and a computation-communication overlap scheduler to collectively address the challenges of compute throughput, memory locality, and communication overheads in PIC simulations.

\subsection{Overall Framework}
POLAR-PIC adopts an \textit{algorithm-memory-communication co-design} philosophy that unifies coupled computation, data organization, and particle migration into a dataflow pipeline. As illustrated in Figure~\ref{fig:framework}, the framework comprises three interdependent modules.

\textbf{First, the compute module reformulates the Field Interpolation for MPUs.} This module processes particles in cell-based batches, transforming the original per-particle inner-product reduction into a 2D outer-product accumulation form. This module relies on an ordered particle layout to prevent irregular gather-load patterns from diminishing MPU throughput.

\textbf{Second, SoW maintains the ordered layout and prepares migrating particles for communication.}
It leverages the particle write-back path to keep resident particles contiguous in memory, while identifying and packing cross-domain particles on the fly, effectively producing the outgoing data needed by redistribution.

\textbf{Third, the redesigned redistribution module overlaps particle communication with Deposition.}
After the Field Interpolation phase, it issues non-blocking \texttt{UNR\_Put} operations to asynchronously transfer the migrant particles packed by SoW. By invoking \texttt{UNR\_Wait} after the Deposition kernel completes, the primary communication latency is effectively hidden within the computation pipeline, leaving only a minimal merge step at the end of the timestep.

In summary, high computational throughput hinges on the ordered in-memory layout maintained by SoW, while SoW also packs migrating particles for the redistribution module to overlap transmission and collection. This co-design enables POLAR-PIC to achieve particle-processing phase performance improvements beyond isolated kernel optimizations on next-generation HPC platforms.

\subsection{Matrixized Field Interpolation}
\label{sec:kernal}

Field Interpolation projects electromagnetic fields from the Eulerian grid onto Lagrangian particle positions and constitutes a typical \textit{gather-stencil} kernel. As illustrated in Figure \ref{fig:gather_tile}, each particle performs a weighted accumulation over $K$ stencil nodes, yielding low operational intensity and an inherent many-to-one inner-product reduction between a weight vector and a field vector. To bridge this structural mismatch with MPU-style outer-product execution, POLAR-PIC introduces an outer-product reformulation via \textbf{Batching through Tensor Stacking}. By logically stacking $N$ particles within the same cell (e.g., $N=8$ in FP64), it casts $N$ independent $1 \times K$ linear contractions into a single $N \times K$ matrix form, thereby shifting the kernel from a 1-D vector pipeline to an MPU-friendly 2-D batched dataflow that exposes outer-product accumulation.
Since typical particle densities far exceed the batch size, full batches are efficiently formed from contiguous SoA segments. Residual particles fewer than $N$ are handled via zero-padding, ensuring a unified matrix execution path.

Let $K$ denote the stencil size for 3-D Interpolation (e.g., $K=4^3=64$ for 3rd-order Interpolation) and  $D$ denote the dimension of field components (typically $D=6$ for $\{E_x,E_y,E_z,B_x,B_y,B_z\}$). For a  particle $p$, the interpolated field $\mathbf{f}_p \in \mathbb{R}^D$ is accumulated from $K$ stencil nodes:
\begin{equation}
\mathbf{f}_p = \sum_{q=1}^{K} w_{p,q} \cdot \mathbf{g}_q,
\label{eq:interp_single}
\end{equation}
where $\mathbf{g}_q \in \mathbb{R}^D$ is the field vector at grid node $q$, and $w_{p,q}$ is the geometric weight scalar for particle $p$ at that node (i.e., the product of shape functions $S_x S_y S_z$).

By stacking $N$ particles from the same cell, we construct a weight matrix $\mathbf{W} \in \mathbb{R}^{N\times K}$ and a grid-field matrix $\mathbf{G} \in \mathbb{R}^{K\times D}$:
\begin{equation}
\mathbf{W} =
\begin{bmatrix}
\mathbf{w}_{p_1}^{\mathsf{T}} \\
\vdots \\
\mathbf{w}_{p_N}^{\mathsf{T}}
\end{bmatrix},
\quad
\mathbf{G} =
\begin{bmatrix}
\mathbf{g}_1^{\mathsf{T}} \\
\vdots \\
\mathbf{g}_K^{\mathsf{T}}
\end{bmatrix},
\label{eq:WG_def}
\end{equation}
where each $\mathbf{w}_{p_i} \in \mathbb{R}^{K}$ stacks the stencil weights of particle $p_i$. The Interpolation result for the entire batch, $\mathbf{F} \in \mathbb{R}^{N \times D}$, can be expressed as the matrix multiplication $\mathbf{F} = \mathbf{W} \mathbf{G}$.
Expanding along the stencil dimension yields a sum of \textit{rank-1 updates}:
\begin{equation}
\mathbf{F} = \sum_{q=1}^{K} \left( \mathbf{w}_{:,q} \otimes \mathbf{g}_q \right),
\label{eq:rank1}
\end{equation}
where $\mathbf{w}_{:,q} \in \mathbb{R}^N$ is the stacked weight vector of $N$ particles for stencil node $q$, and $\mathbf{g}_q \in \mathbb{R}^D$ represents the corresponding field-component vector. This form maps directly to the MPU dataflow. The loop iterates over stencil nodes, loads one $\mathbf{g}_q$, and accumulates $\mathbf{w}_{:,q}\otimes\mathbf{g}_q$ into a tile register via MOPA. For a field vector with $D=6$, $\mathbf{g}_q$ can be zero-padded to the tile width (e.g., 8), leading to the tile update:
\begin{equation}
\mathbf{F}_{\text{tile}} \leftarrow \mathbf{F}_{\text{tile}} + \mathbf{w}_{:,q} \otimes \mathbf{g}_q.
\label{eq:tile_update}
\end{equation}

\subsection{Sort-on-Write Algorithm}

The performance of MPU-based Interpolation hinges on contiguous data supply, which traditional index-based sorting~\cite{rao2026matrix} fails to guarantee. To resolve the memory fragmentation bottleneck, we propose the \textbf{Sort-on-Write (SoW)} mechanism. As shown in Figure~\ref{fig:gather_tile} and Algorithm~\ref{alg:sow_full_flow}, SoW exploits the \textit{read-modify-write} dataflow of particle push to transform explicit reordering into an \textbf{inlined streaming split}. This design dynamically maintains a physically contiguous layout without a periodic stop-the-world global sort.

\begin{algorithm}[h]
\caption{SoW-Enabled Time-Step: Stream-Split, Comm-Fusion, and Layout Reuse}
\label{alg:sow_full_flow}
\begin{algorithmic}[1]
\Require Particles $\mathcal{P}_{cur}$ (Ordered + Indexed), Fields $\mathbf{E}$, $\mathbf{B}$, Comm Buffers $\mathcal{B}_{local}$, $\mathcal{B}_{remote}$, Pre-defined Masks $\mathbf{m}_{stay}$, $\mathbf{m}_{move}$, $\mathbf{m}_{rmt}$
\Ensure Updated $\mathcal{P}_{next}$ partitioned for reuse; Comm initiated; Current/Charge $\mathbf{J}$ deposited

\State \textit{// Disordered particles in $\mathcal{P}_{cur}$ are pre-binned for logical traversal}
\State Initialize cursors: $ptr_{ord} \leftarrow 0$, $ptr_{dis} \leftarrow \text{Capacity}(\mathcal{P}_{next})$;

\Statex \textit{// Phase 1: Interpolation \& Push with SoW}
\For{each cell $c \in \text{Tile}\ T$}
    \State $\text{Meta}[c].\text{start} \leftarrow ptr_{ord}$; \Comment{start offset for next frame}
    \State $\mathcal{S}_{ord} \leftarrow \text{\textbf{VPU\_Load}}(\mathcal{P}_{cur}.\text{Ordered}[c])$;
    \State $\mathcal{S}_{idx} \leftarrow \text{\textbf{VPU\_GatherLoad}}(\mathcal{P}_{cur}.\text{Disordered}, \text{bin\_to\_ip}[c])$;
    
    \For{batch $\mathbf{P}_{in} \in \text{\hskip2em}(\mathcal{S}_{ord}, \mathcal{S}_{idx})$} 
        \State $\mathbf{P}_{new} \leftarrow \textbf{MPU\_InterpolationKernel}(\mathbf{P}_{in}, \mathbf{E}, \mathbf{B})$ ;
        
        \Statex \hskip3em \textit{// In-register Classification \& Stream-Split}
        \State $\mathbf{c}_{new} \leftarrow \text{\textbf{GetCellID}}(\mathbf{P}_{new})$;
        \State $\mathbf{m}_{stay} \leftarrow (\mathbf{c}_{new} == c)$; 
        \State $\mathbf{m}_{move} \leftarrow (\neg \mathbf{m}_{stay})$;
        \Statex \hskip3em \textit{// Keep Resident physically contiguous}
        \State \textbf{CompactStore}($\mathcal{P}_{next}[ptr_{ord}]$, $\mathbf{P}_{new}$, $\mathbf{m}_{stay}$) ;
        \State $ptr_{ord} \leftarrow ptr_{ord} + \text{\textbf{PopCount}}(\mathbf{m}_{stay})$;
        
        \If{$\text{Any}(\mathbf{m}_{move})$}
            \State $N_{move} \leftarrow \text{\textbf{PopCount}}(\mathbf{m}_{move})$;
            \Statex \hskip4em \textit{// Grow Disordered Region from Tail}
            \State $ptr_{dis} \leftarrow ptr_{dis} - N_{move}$;
            \State \textbf{CompactStore}($\mathcal{P}_{next}[ptr_{dis}]$, $\mathbf{P}_{new}$, $\mathbf{m}_{move}$);
            
            \Statex \hskip4em \textit{// Fusion: Pre-pack based on Destination}
            \State $\mathbf{m}_{rmt} \leftarrow \text{\textbf{IsRemoteRank}}(\mathbf{P}_{new}, \mathbf{m}_{move})$;
            \State \textbf{CopyToBuffer}($\mathcal{B}_{local}$, $\mathbf{P}_{new}$, $\mathbf{m}_{move} \land \neg \mathbf{m}_{rmt}$);
            \State \textbf{CopyToBuffer}($\mathcal{B}_{remote}$, $\mathbf{P}_{new}$, $\mathbf{m}_{rmt}$);
        \EndIf
    \EndFor
    \State $\text{Meta}[c].\text{length} \leftarrow ptr_{ord} - \text{Meta}[c].\text{start}$
\EndFor
\State \textbf{TriggerUNR\_BatchPut}($\mathcal{B}_{remote}$, $\text{PopCount}(\mathbf{m}_{rmt})$);

\Statex \textit{// Phase 2: Deposition with Layout Reuse}
\For{each cell $c \in \text{Tile} \ T$} \Comment{Reuse ordered layout}
    \State $\mathbf{P}_{ord} \leftarrow \text{VPU\_Load}(\mathcal{P}_{next}, \text{Meta}[c].\text{start}, \text{Meta}[c].\text{length})$
    \State \textbf{MPU\_DepositKernel}($\mathbf{P}_{ord}, \mathbf{J}$) 
\EndFor

\For{particle $p \in \text{Range}(ptr_{dis}, \text{Capacity})$} \Comment{Fallback to VPU}
    \State \textbf{VPU\_DepositKernel}($p, \mathbf{J}$) 
\EndFor

\end{algorithmic}
\end{algorithm}

\subsubsection{Dual-Region Memory Management.}

Leveraging the CFL condition that most particles remain local within a time step, we partition each tile's memory into a dominant \textbf{Ordered Region} (physically contiguous) and a small \textbf{Disordered Region} (append-only tail). This dual-region layout is maintained using two pre-allocated buffers per rank-local tile, sized by a runtime upper-bound heuristic to avoid reallocation during timesteps while keeping the extra space cost local to each tile. Runtime pointer swapping reuses these buffers across timesteps, thereby eliminating frequent dynamic allocation overheads without resorting to full-volume global double-buffering. Within this layout, the Ordered Region supports direct MPU vector loading, whereas the Disordered Region relies on a lightweight auxiliary index (\texttt{bin\_to\_ip}) to recover logical traversal order. This hybrid organization keeps the vast majority of memory accesses contiguous while confining fragmentation to a small tail. During the SoW process, those disordered particles are progressively absorbed back into the Ordered Region of the next buffer, naturally restoring physical locality over time.

\subsubsection{Execution Flow and Layout Reuse.}

The SoW execution follows a fused pipeline (Algorithm~\ref{alg:sow_full_flow}):
\begin{itemize}[leftmargin=10pt, labelsep=0.5em]
    \item \textbf{Tail Sorting:} At the beginning of each time step, we perform a low-cost $O(N_{tail})$ binning pass on the Disordered Region to enable logical-order traversal. 
    \item \textbf{Stream-Split Write-back (Lines 9-22):} During the update phase, SIMD instructions classify particles as \textit{resident} or \textit{migrating}. A dual-pointer streaming store then separates them: resident particles are compacted into the Ordered Region of the next buffer, while migrating particles are appended to the Disordered tail.
\end{itemize}
This mechanism acts as a self-healing process, continuously absorbing disordered particles back into the ordered layout. Furthermore, since position updates preserve cell indices for the duration of the step, the \textbf{Deposition} kernel (Lines 26-30) directly reuses this contiguous layout for high-throughput MOPA execution, falling back to VPU atomics only for the sparse tail.

\subsection{Overlapping Particle Communication with Deposition}

\begin{figure}[h]
\centering
\includegraphics[width=\linewidth]{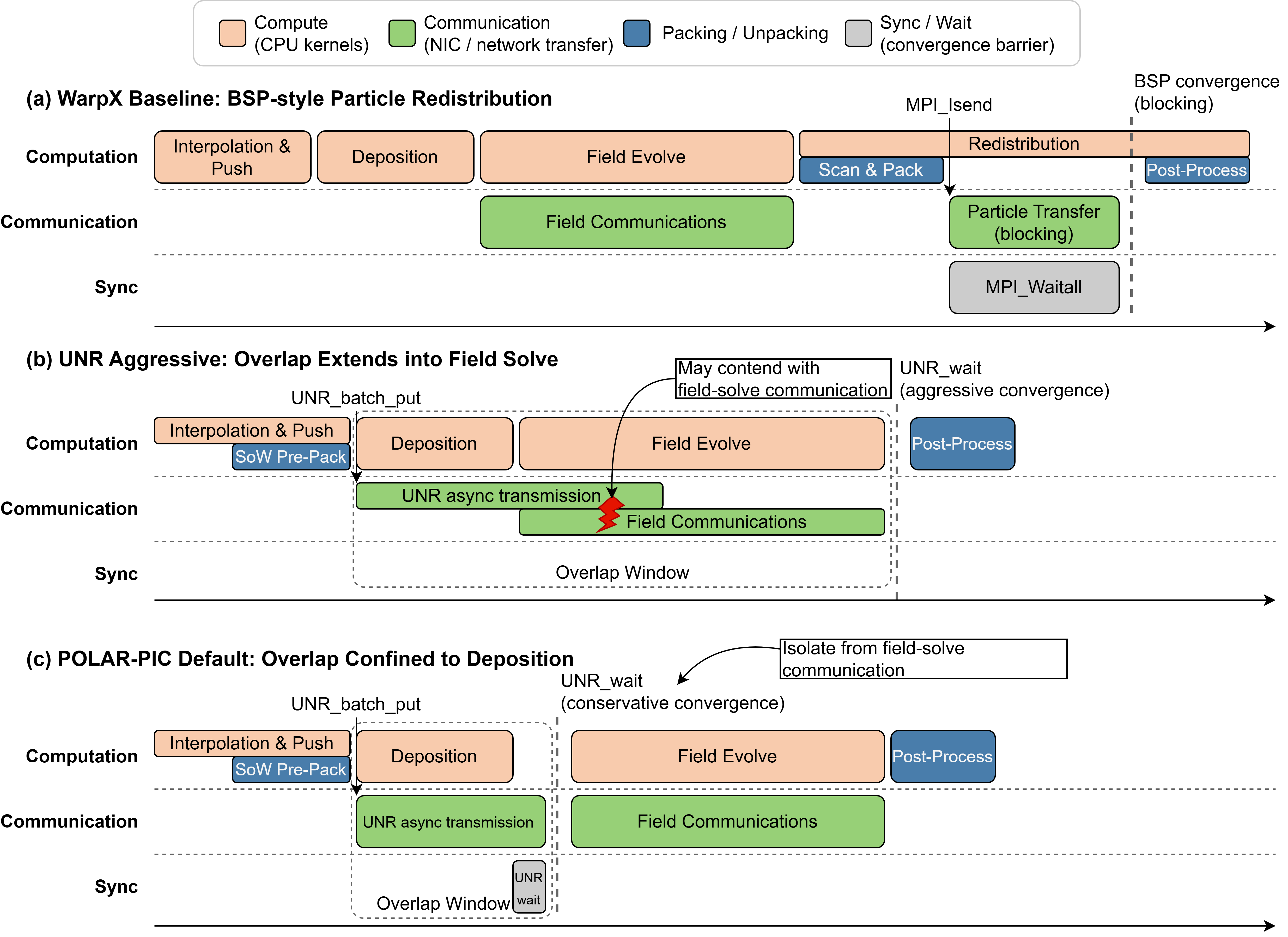}
\caption{\textbf{Comparison of Particle Redistribution Strategies.} }
\label{fig:overlap_pipeline}
\end{figure}

From a dataflow perspective, the updated particle position $\mathbf{x}^{n+1}$ is available at the end of \textit{Particle Push}, and subsequent Deposition and field update do not change the particle's spatial mapping to grid indices. Accordingly, redistribution can be partially initiated early by (i) determining the destination tile/rank from $\mathbf{x}^{n+1}$ and (ii) pre-packing a copy of migrating particle attributes into pre-allocated remote send buffers or local tiles. As illustrated in Figure~\ref{fig:overlap_pipeline}, this design transforms the traditional blocking communication phase into an asynchronous pipeline, effectively masking transmission latency while avoiding interference with field solvers.

\subsubsection{Re-packing via Kernel Fusion}
\label{sec:repack}
In native WarpX redistribution, migrant particles are identified and packed through a separate end-of-step scan, which introduces redundant traversal and additional memory traffic. POLAR-PIC removes this standalone scan-and-pack stage by fusing pre-packing into the write-back path, where SoW already exposes the destination information of migrated particles. The runtime precomputes neighbor connectivity under the static domain decomposition and registers communication buffers during initialization; thread-private send/recv buffers are used to reduce synchronization and avoid consolidating per-thread outputs.

During write-back, migrants are dispatched directly to their destinations. Intra-rank migrants are written from registers into the target tile's receive buffer, while inter-rank migrants are packed into the thread-private UNR send buffer for the corresponding neighbor. Meanwhile, particles leaving the current tile are swapped into the tail disordered region so they can be removed efficiently at the end of the step. Buffers follow a linear \texttt{[Header | Payload]} layout with lightweight per-thread offsets, and UNR batch transmission avoids extra process-level \texttt{memcpy} and cache pollution. Importantly, this design does not imply that all packing work disappears; rather, the remaining packing work is absorbed into the write-back path instead of appearing as a separate end-of-step stage.

\subsubsection{Asynchronous Transmission with UNR}
After pre-packing completes, the main thread invokes UNR \texttt{batch\_put} outside the OpenMP region to submit all RDMA writes across threads and neighbor directions in one batch. The NIC then transfers data from thread-local send buffers into remote receive buffers, while the CPU proceeds immediately to the Deposition kernel, thereby overlapping communication with Deposition. A receiver process only waits for completion signals via \texttt{UNR\_Wait} to detect data arrival, rather than posting receive-side synchronization operations such as \texttt{MPI\_Recv}, \texttt{MPI\_Irecv/Wait}, or \texttt{MPI\_Win\_Fence}.

\subsubsection{Convergence and Particle Finalization}
POLAR-PIC confines overlap to the Deposition window and enforces convergence with \texttt{UNR\_Wait} immediately after Deposition to avoid interference with the latency-sensitive field-solve communication. This separation reduces contention in NIC queues and network injection bandwidth and improves timestep stability; in practice, most particle transfers complete before Deposition ends, making the post-Deposition wait negligible.

After convergence, received particles are unpacked and finalized. Since SoW clusters invalid entries at the tail disordered region, deletion reduces to tail truncation by updating the SoA size pointer. Incoming \texttt{[Header | Payload]} segments are then unpacked and appended to the target tile’s SoA tail using parallel \texttt{memcpy}. Overall, POLAR-PIC refactors redistribution from an end-of-step blocking BSP phase into a pipelined overlap scheme that achieves effective masking while isolating bandwidth interference.

\section{Experimental Setup}

\subsection{Experimental Platform}

\begin{figure}[h]
  \centering
  \includegraphics[width=0.9\linewidth]{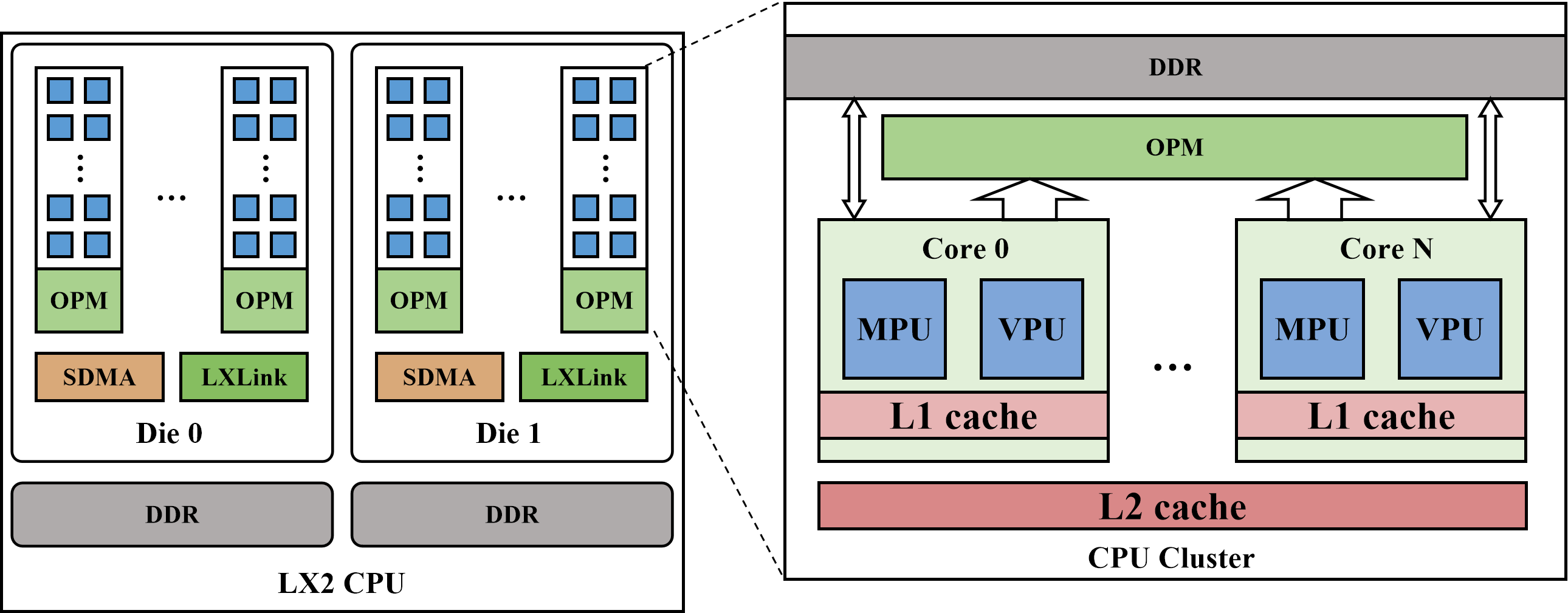} 
  \caption{Architectural diagram of the LX2 processor, illustrating the dual-Die design, core distribution, and key components like the VPU, MPU, and memory interfaces.}
  \label{fig:lx2_arch}
\end{figure}

Our primary experiments were conducted on the LS pilot system, a next-generation HPC cluster. Each compute node incorporates two LX2 high-performance CPUs. As illustrated in the architectural diagram in Figure~\ref{fig:lx2_arch}, each processor package contains over 256 cores distributed across two compute dies. These cores support both vector (VPU) and matrix (MPU) computation engines. The VPUs execute double-precision (FP64) SIMD instructions, while the MPUs are designed for 8$\times$8 matrix operations. Critically, the MPU's MOPA instruction offers approximately 4$\times$ the theoretical FP64 performance of the VPU's Multiply-Accumulate (MLA) instruction, presenting a significant opportunity for acceleration. Both compute units operate at frequencies at or above 1.3 GHz.

The LX2 CPU implements a system-on-chip design where each Die features 128GB of off-die DDR memory organized across 4 NUMA domains. The Dies are interconnected through an LXLink network, which provides up to 48\,GB/s bidirectional bandwidth per NIC and supports RDMA operations. The LS system employs a customized Linux-based OS with LLVM-based Clang/Flang compilers. All binaries are compiled with \texttt{-O3} and \texttt{-flto} and linked against an optimized OpenMPI library and architecture-tuned math kernels. For one-sided transfers, we utilize the latest UNR library tuned for LS RDMA verbs. 
For comparison, the MPI-based redistribution is implemented using non-blocking \texttt{MPI\_Isend}/\texttt{MPI\_Irecv} primitives to provide asynchronous progress.

For cross-platform comparison, we conducted experiments on a platform equipped with two 28-core CPUs and 8 NVIDIA A800 GPUs, which features 80 GB of HBM2e memory. The software environment used for these tests included NVIDIA driver version 535.104.12.

\subsection{Workloads and Configuration}

We implement the \textbf{POLAR-PIC} framework atop the widely adopted \texttt{WarpX v24.07}~\cite{fedeli2022warpx}. While our current implementation targets WarpX, the proposed architectural optimizations are generic and portable to other standard PIC workflows. 
Performance comparisons and ablation studies are conducted on 2 compute nodes using a hybrid MPI+OpenMP parallelization strategy. We configure one MPI rank per NUMA domain (aggregating 16 ranks per node), spawning 32 OpenMP threads per rank with strict thread affinity enforced to minimize context switching overhead. For large-scale weak scalability experiments, we utilize the Uniform Plasma benchmark on LS pilot system, scaling up to 4,096 nodes with the same per-node configuration.

We disable I/O to isolate computational performance. All simulations employ a 3D Cartesian grid with 3rd-order B-spline shape factors, the Yee field solver, direct current Deposition, and the Boris particle pusher. We execute 100 timesteps for measurement after an initial warm-up phase. We use two workloads: (i) a \textbf{Uniform Plasma} microbenchmark with a grid of $256 \times 128 \times 128$, where we sweep the particle density, i.e., particles per cell (PPC), to vary computational intensity and memory pressure, and adjust the thermal velocity $u_{th}$ to induce different levels of particle migration; and (ii) a \textbf{Laser-Ion Acceleration with a Planar Target} production case with a grid of $192 \times 192 \times 256$ over an approximately $7.5\,\mu m \times 7.5\,\mu m \times 15\,\mu m$ domain to validate particle-phase performance under strongly non-uniform particle distributions and significant migration. All experiments are repeated three times excluding warm-up overhead; we report the average execution time across these trials to filter out transient system jitter. Detailed physical and numerical parameters are listed in Appendix~\ref{append:parameter}.

\subsection{Evaluation Metrics}

We quantify performance using a multi-level metric suite. To isolate the architectural impact on particle-grid interactions, we define overall \textbf{Particle-Phase Time} as $T_{particle} = T_{Interpolation} + T_{deposit} + T_{redistribute}$, where $T_{Interpolation}$ includes Field Interpolation, Push, SoW, and fused re-packing overheads; $T_{deposit}$ includes current Deposition overhead; and $T_{redistribute}$ captures particle communication issuing, explicit waiting, and post-processing. Furthermore, $T_{steps} = T_{particle}/N_{steps}$ denotes the \textbf{average particle-phase time per step}. Unless otherwise specified, all derivative metrics in this work (including speedups, throughput, and efficiency) are normalized against $T_{particle}$ or $T_{steps}$ to exclude the Eulerian field solver and global halo exchanges, which remain identical to the native WarpX reference pipeline on LX2. Our primary optimization target is therefore the particle-processing critical path, while end-to-end full-timestep behavior is reported separately in the weak-scaling study.

For ablations, we further decompose computation into $T_{prep}$ (index / shape-factor preparation), $T_{sort}$ (IncrSort or SoW overhead), $T_{kernel}$ (pure math execution), and $T_{reduce}$ (write-back / reduction); 
and we \textbf{decompose communication} into $T_{pack}$ (serialization and packing), $T_{issue}$ (issuing MPI / UNR calls), $T_{wait}$ (explicit synchronization), and $T_{post\_process}$ (unpacking, appending, and invalid-particle removal). 

To evaluate particle processing throughput and computational density, we report \textbf{Particles per Second} (PPS, $N_{total} / T_{steps}$) and \textbf{Cycles per Particle} (CPP), the latter being normalized to a 1.3 GHz reference frequency. The \textbf{Effective Transport Overlap Ratio} $\eta_{\text{overlap}}$ is defined as:

\begin{equation}
\eta_{\text{overlap}} = 1 - \frac{T_{issue}^{\text{Overlap}}+T_{wait}^{\text{Overlap}}}{T_{issue}^{\text{baseline}} + T_{wait}^{\text{baseline}}}
\end{equation}

where the numerator represents the explicit communication time under overlap, and the denominator is the exposed particle-communication time of the BSP-style native \texttt{WarpX} redistribution path. Since the BSP approach prohibits overlap, $\eta_{\text{overlap}}$ directly quantifies the fraction of communication latency removed from the critical path relative to synchronous execution. 
In the results, we discuss residual synchronization through the communication-critical path and report a representative maximum per-rank wait time

Finally, we evaluate hardware utilization and particle-phase quality. \textbf{Peak efficiency} is defined as:
\begin{equation}
\eta_{\text{peak}} = \max\left(\frac{\sum \text{Particle FLOPs}}{T_{steps} \cdot P_{\text{theoretical}}}\right) \times 100\%,
\end{equation}
where $P_{\text{theoretical}}$ is the theoretical FP64 peak of the platform.
Particle FLOPs are standardized using the native WarpX implementation, counting 1,636 and 419 FLOPs per particle for the Interpolation and Deposition kernels, respectively.

Following standard PIC community practice~\cite{fedeli2022warpx}, we also report the \textbf{per-node Figure of Merit} ($\text{FOM}_{\text{node}}$):
\begin{equation}
\text{FOM}_{\text{node}} = \max\left(\frac{\alpha N_c + \beta N_p}{T_{steps} \cdot N_{nodes}}\right),
\end{equation}
where $N_c$ and $N_p$ are the total number of grid cells and macroparticles, respectively. The weights $\alpha=0.1$ and $\beta=0.9$ are adopted to ensure consistency with WarpX benchmarks~\cite{fedeli2022warpx}.

\subsection{Experimental Design}

Table~\ref{tab:exp_config} summarizes the ablation variants and integrated system configurations. Here C0 corresponds to the native BSP-style WarpX particle-redistribution path without communication-computation overlap. For sorting-based comparisons, Matrix-PIC serves as the state-of-the-art reference, representing the logical-order optimization path beyond the native WarpX baseline. We bold the default POLAR-PIC setup, and provide full implementation details in Appendix~\ref{append:implementation}.

\begin{table}[h]
\centering
\caption{Summary of experimental configurations and ablation variants. }
\label{tab:exp_config}
\resizebox{\columnwidth}{!}{%
\begin{tabular}{l l}
\toprule
\textbf{Variant} & \textbf{Key Configuration Characteristics} \\
\midrule
\multicolumn{2}{l}{\textit{\textbf{Exp 1: Interpolation (Gather \& Push) Variants}}} \\
G0 (Baseline) & WarpX native, Compiler auto-vectorization (VPU), Unsorted \\
G1 & Hand-tuned VPU intrinsics, Unsorted \\
G2 & VPU, Logical Index-Sort (Reproduces Matrix-PIC~\cite{rao2026matrix}) \\
G3 & VPU, Physical Reordering (based on G2 indices) \\
G4 & VPU, Sort-on-Write (SoW) Physical Reordering \\
G5 & MPU, Logical Index-Sort (MPU version of G2) \\
G6 & MPU, Physical Reordering (MPU version of G3) \\
\textbf{G7 (POLAR-PIC)} & \textbf{MPU, SoW Physical Reordering} \\
\midrule
\multicolumn{2}{l}{\textit{\textbf{Exp 2: Deposition (Scatter) Variants}}} \\
D0 (Baseline) & WarpX native, Compiler auto-vectorization (VPU), Unsorted \\
D1 & MPU, Reuse G2's Logical Index (Reproduces Matrix-PIC~\cite{rao2026matrix}) \\
D2 & MPU, Reuse G7's Physical Layout + Tail Binning \\
\textbf{D3 (POLAR-PIC)} & \textbf{MPU, Reuse G7's Physical Layout + VPU Tail} \\
\midrule
\multicolumn{2}{l}{\textit{\textbf{Exp 3: Overlap Strategy Variants}}} \\
C0 (Baseline) & No Overlap (BSP-style blocking redistribution) \\
C1 & MPI-based, Conservative sync (Post-Deposit) \\
\textbf{C2 (POLAR-PIC)} & \textbf{UNR-based, Conservative sync (Post-Deposit)} \\
C3 & MPI-based, Aggressive sync (Post-FieldSolve) \\
C4 & UNR-based, Aggressive sync (Post-FieldSolve) \\
\midrule
\multicolumn{2}{l}{\textit{\textbf{Integrated Systems}}} \\
WarpX-Native (Baseline) & Native \texttt{WarpX v24.07} reference pipeline on LX2 (G0 + D0 + C0) \\
Matrix-PIC & State-of-the-art logical-order pipeline (G2 + D1 + C0) \\
\textbf{POLAR-PIC} & \textbf{Full Outer-Product Pipeline (G7 + D3 + C2)} \\
\bottomrule
\end{tabular}%
}
\end{table}

\section{Experimental Results}

\subsection{Overall Performance}

Unless otherwise stated, the results in this subsection focus on particle-phase performance, which is the primary optimization target of POLAR-PIC. End-to-end full-timestep behavior is reported separately in the weak-scaling study in Section\ref{sec-exp:scale}.

\subsubsection{Uniform Plasma Microbenchmarks}

We first evaluate peak particle throughput and dynamic robustness by sweeping particle density (PPC) and thermal velocity ($u_{th}$). Relative to the native WarpX reference pipeline on LX2, \textbf{POLAR-PIC} achieves up to $10.9\times$ particle-phase speedup in high-density regimes and still delivers a $7.3\times$ advantage under high migration intensity; it also outperforms Matrix-PIC by up to $4.7\times$.

\begin{figure}[h]
\centering
\includegraphics[width=\linewidth]{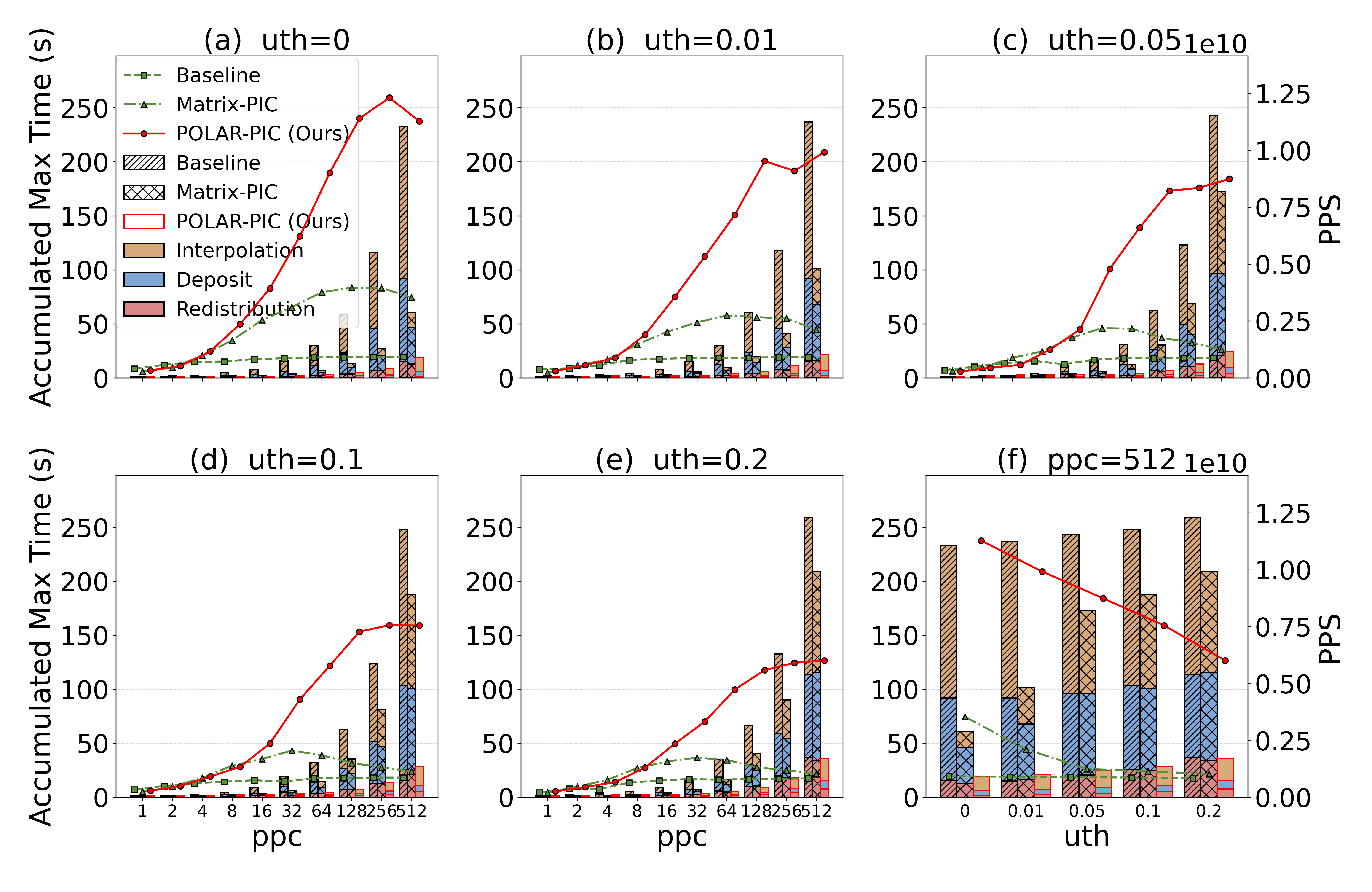}
\caption{Uniform-plasma particle-phase performance under varying particle density (PPC) and thermal velocity $u_{th}$. }
\label{fig:uniform_perf}
\end{figure}

As illustrated in Figure~\ref{fig:uniform_perf}, the WarpX-Native reference is bounded by the throughput ceiling of the VPU path. While Matrix-PIC delivers significant speedups at moderate densities, its performance degrades at high PPC ($>256$) due to escalating index-maintenance overheads. \textbf{POLAR-PIC} avoids this scalability pitfall by relying on the SoW mechanism, preserves physical contiguity and yields a near-monotonic throughput trend as PPC increases. At very low densities ($PPC=1$), speedups naturally diminish for matrix-based variants because padding and metadata costs are hard to amortize when batches are underfilled, consistent with prior observations on low-occupancy matrix kernels~\cite{rao2026matrix}.

\begin{table}[b]
\centering
\caption{Summary of $T_{Particle}$, PPS, CPP in Uniform Plasma at representative PPC points ($u_{th}=0.01$).}
\label{tab:uniform_summary}
\resizebox{\columnwidth}{!}{%
\begin{tabular}{c l r r r r}
\toprule
PPC & Scheme & $T_{Particle}$ (s) & PPS (Gparticles/s) & CPP (cycles/particle) & Speedup ($\times$) \\
\midrule
\multirow{3}{*}{1}     & \textbf{Baseline (WarpX)} & \textbf{1.102}  & \textbf{0.381} & \textbf{3.416} & \textbf{1.00} \\
                       & Matrix-PIC       & 1.852  & 0.226 & 5.741 & 0.60 \\
                       & POLAR-PIC        & 1.379  & 0.304 & 4.275 & 0.80 \\
\cmidrule(l){1-6}
\multirow{3}{*}{8}     & Baseline (WarpX) & 4.278  & 0.784 & 1.657 & 1.00 \\
                       & Matrix-PIC       & 2.290  & 1.465 & 0.887 & 1.87 \\
                       & \textbf{POLAR-PIC}        & \textbf{1.769}  & \textbf{1.896} & \textbf{0.686} & \textbf{2.42} \\
\cmidrule(l){1-6}
\multirow{3}{*}{64}    & Baseline (WarpX) & 30.280 & 0.887 & 1.466 & 1.00 \\
                       & Matrix-PIC       & 9.798  & 2.740 & 0.475 & 3.09 \\
                       & \textbf{POLAR-PIC}        & \textbf{3.750}  & \textbf{7.158} & \textbf{0.182} & \textbf{8.07} \\
\cmidrule(l){1-6}
\multirow{3}{*}{512}   & Baseline (WarpX) & 236.939 & 0.906 & 1.434 & 1.00 \\
                       & Matrix-PIC       & 101.673 & 2.112 & 0.615 & 2.33 \\
                       & \textbf{POLAR-PIC}        & \textbf{21.651}  & \textbf{9.919} & \textbf{0.131} & \textbf{10.94} \\
\bottomrule
\end{tabular}%
}
\end{table}

When sweeping thermal velocity ($u_{th}$) to emulate dynamic disorder, the advantages of our physical layout management become even more pronounced. While the Baseline shows limited sensitivity to disorder because it does not exploit locality, sorting-based methods face challenges as disorder increases. Specifically, at $u_{th}=0.2$, Matrix-PIC’s speedup collapses to just $1.2\times$ due to severe data supply stalls. Conversely, POLAR-PIC demonstrates superior resilience via the SoW mechanism. Even under this worst-case turbulence, it sustains a $7.3\times$ speedup over the Baseline, maintaining contiguous SoA segments for the MPU where logical ordering fails.

\subsubsection{Laser-Ion Acceleration Benchmark}
In the highly non-uniform LIA benchmark, \textbf{POLAR-PIC} sustains a robust $4.4\times$ particle-processing speedup over the Baseline and $3.8\times$ over Matrix-PIC, significantly outperforming logical-ordering alternatives in migration-heavy scenarios.

\begin{figure}[h]
\centering
\includegraphics[width=\linewidth]{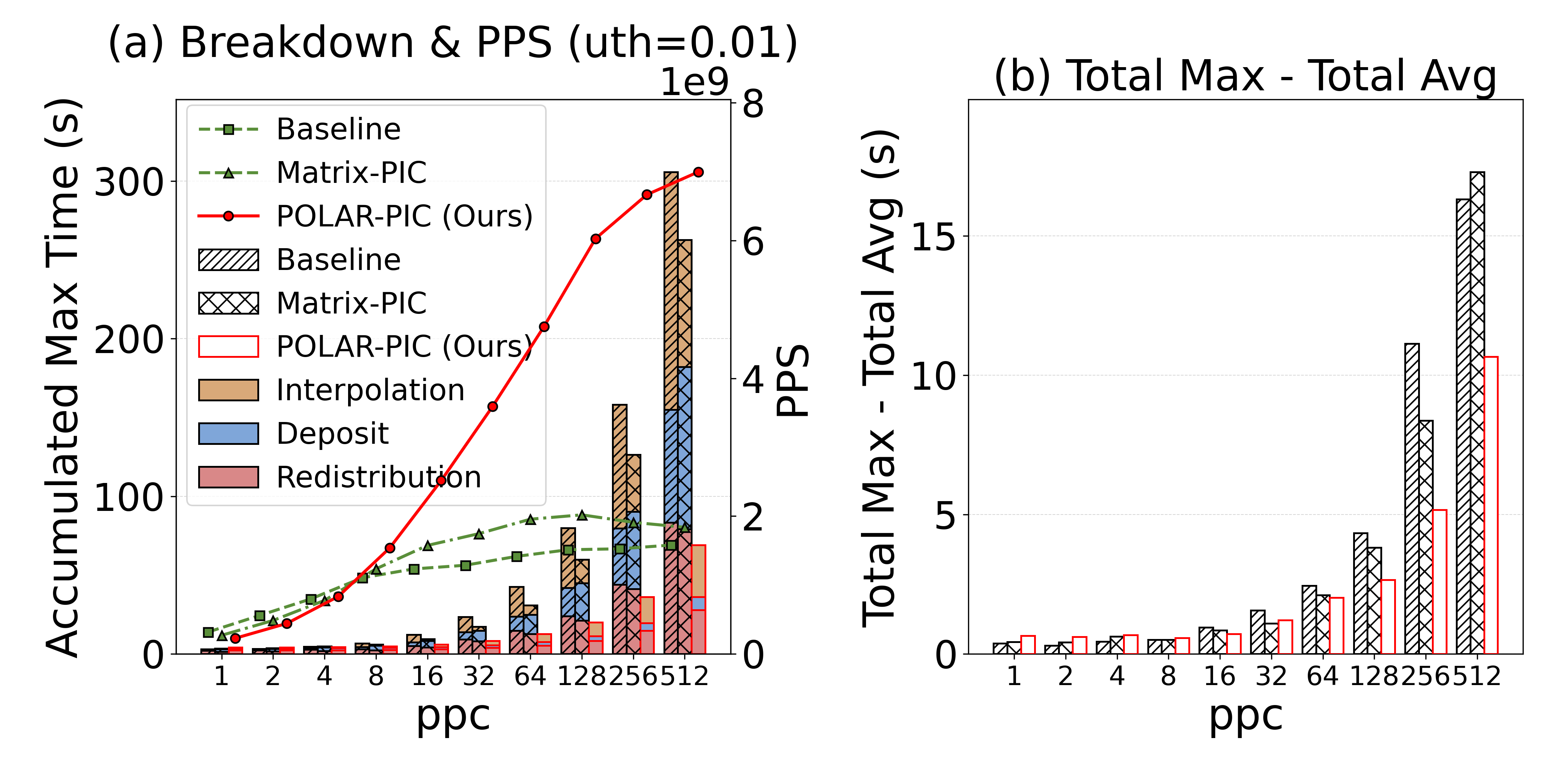}
\caption{Particle-Phase Performance and Long-tail Effect Analysis of Laser-Ion Acceleration Simulations with Varying PPC.}
\label{fig:laser_ion_perf}
\end{figure}

As shown in Figure~\ref{fig:laser_ion_perf}, the performance advantage stems primarily from the optimized handling of particle redistribution, which emerges as the dominant cost in this regime. POLAR-PIC addresses this bottleneck through its asynchronous pipeline, accelerating the redistribution phase by $3.0\times$ relative to the Baseline. Crucially, the analysis of long-tail latency highlights a major stability gap. 
Matrix-PIC suffers from severe performance jitter due to frequent index rebuilding, with the (max $-$ mean) deviation exceeding that of the Baseline by $6.1\%$. In contrast, by leveraging UNR-based NIC offloading and computation-communication overlap, POLAR-PIC effectively masks synchronization overheads. This design reduces the cumulative long-tail latency by $34.6\%$ compared to the Baseline and $38.3\%$ compared to Matrix-PIC, confirming that POLAR-PIC delivers not just higher throughput, but predictable stability essential for highly dynamic production simulations.


\begin{figure}[h]
\centering
\includegraphics[width=\linewidth]{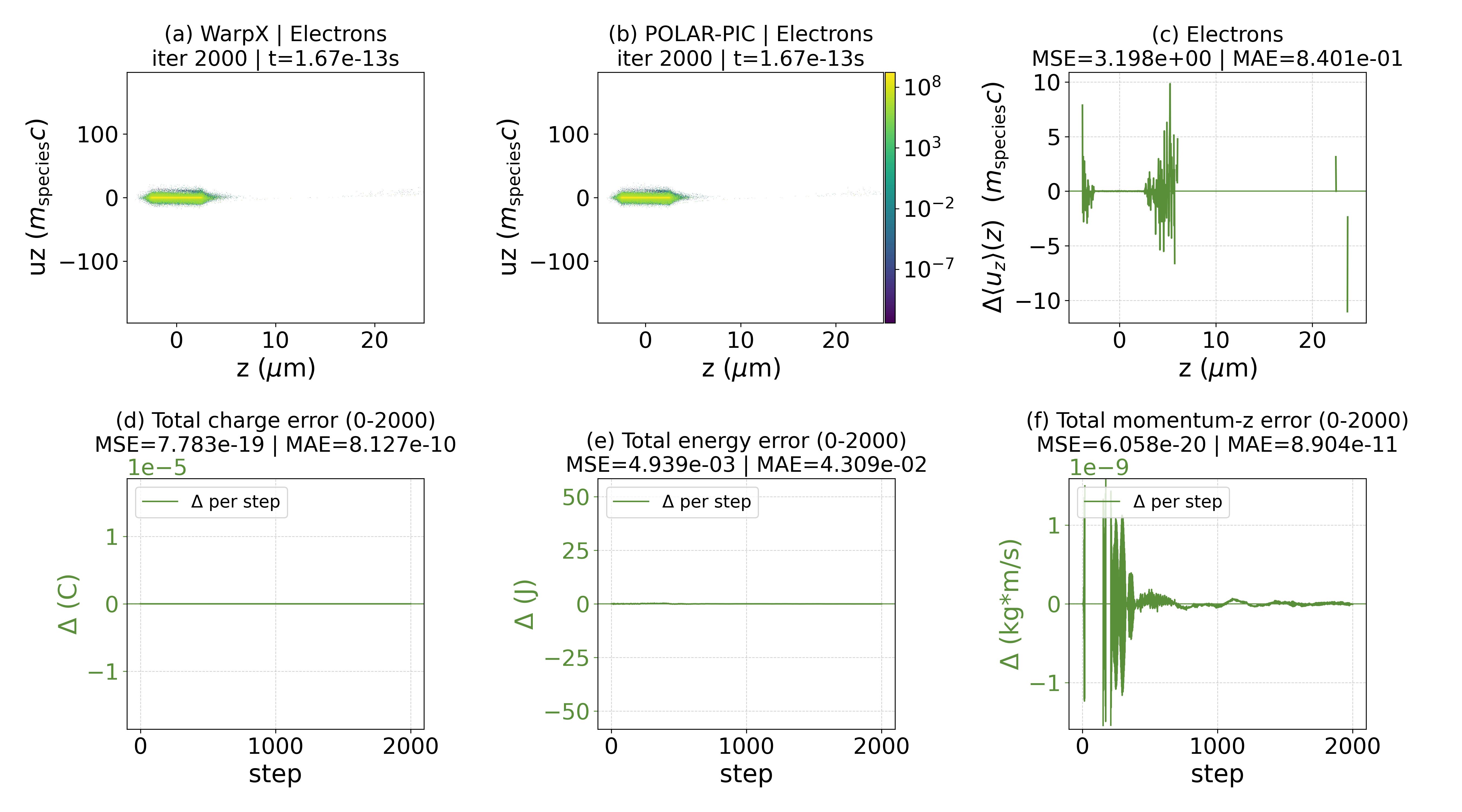}
\caption{Correctness verification for Laser-Ion Acceleration at step 2000 and over the full 2000-step evolution. }
\label{fig:verification_lia}
\end{figure}

\subsubsection{Correctness Verification for Laser-Ion Acceleration}

To ensure numerical fidelity, we conducted a controlled comparison against the native WarpX baseline on the LX2 platform using the LIA benchmark. Figure~\ref{fig:verification_lia} reports a qualitative phase-space comparison at step 2000, together with the per-step errors of total charge, total energy, and longitudinal momentum over 2000 steps. As illustrated in Figure~\ref{fig:verification_lia}, the electron longitudinal phase-space structures and beam morphologies remain highly consistent between native WarpX and POLAR-PIC. Quantitative phase-space assessment yields a Mean Squared Error (MSE) of $3.198$ and a Mean Absolute Error (MAE) of $0.8401$ (normalized to $m_e c$). Given that the peak momentum exceeds $100\,m_e c$, the MAE corresponds to a relative error below $0.84\%$. The remaining discrepancies are primarily attributed to the altered floating-point summation order inherent in the matrix outer-product reformulation. Meanwhile, the per-step error curves in Figure~\ref{fig:verification_lia}(d--f) show that the total charge, total energy, and longitudinal momentum errors remain stably bounded throughout the 2000-step evolution. Together, these results indicate that POLAR-PIC preserves numerical consistency without introducing visible artifacts or instability in this benchmark.

\subsection{Ablation 1: Interpolation Kernel Evolution}

In these experiments, we fix the Deposition strategy to the native WarpX deposition path (D0) and disable communication overlap to isolate the impact of data supply strategies and kernel formulations.


\subsubsection{Stability of SoW Data Supply (G0--G4)}

We first evaluate the stability of data supply strategies under varying migration intensities, finding that the SoW mechanism (G4) sustains high throughput ($5.2\times10^9$ PPS) even at $u_{th}=0.2$.

\begin{figure}[h]
\centering
\includegraphics[width=\linewidth]{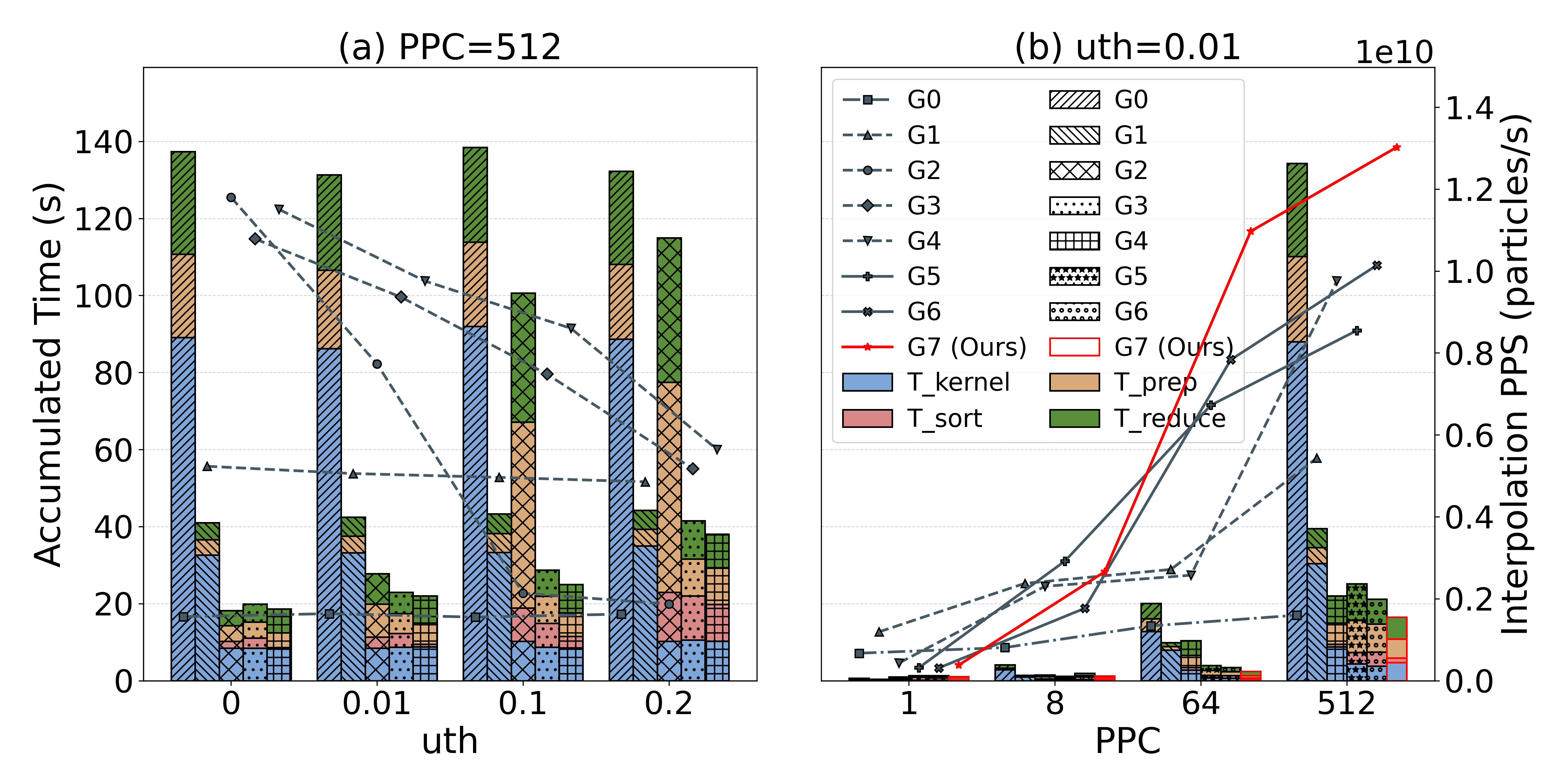}
\caption{VPU results for G0--G4 at $PPC=512$, and MPU Interpolation results for G5--G7 across PPC at $u_{th}=0.01$ (with G0/G1 shown as baselines).}
\label{fig:ablation_a_sve}
\end{figure}

As shown in Figure~\ref{fig:ablation_a_sve}(a), while unsorted baselines (G0/G1) are insensitive to dynamics due to their low-throughput slow sequential access, logical ordering (G2) provides benefits only in low-migration regimes ($u_{th} < 0.1$). As disorder intensifies, the indirect memory accesses in G2 incur escalating address-generation overheads. Specifically,$T_{prep}$ and $T_{reduce}$ increase by $5.6\times$ as $u_{th}$ rises from 0.01 to 0.1, neutralizing the benefits of logical grouping. Conversely, G3 and G4 demonstrate that physical continuity is critical for sustaining bandwidth. Although G3 improves locality via index-guided write-back, its sorting overhead is $1.19\times$ higher than G4's tail-sorting approach at $u_{th}=0.2$. By implementing the dual-region SoW mechanism, G4 eliminates redundant sorting for resident particles, maintaining optimal data supply bandwidth across all dynamic regimes.

\subsubsection{Impact of Data Supply on MPU Interpolation (G0--G1, G5--G7)}

Fixing migration at $u_{th}=0.01$, we sweep particle density to demonstrate that physically contiguous data supply is the prerequisite for unlocking the full acceleration potential of the matrix outer-product operator.

\begin{table}[h]
\centering
\caption{Performance summary for Interpolation (G0--G7) and Deposition (D0--D3) variants under representative conditions (PPC=512, $u_{th}=0.01$)}
\label{tab:ablation_combined_summary}
\resizebox{\columnwidth}{!}{%
\begin{tabular}{l r r r r r}
\toprule
Variant & Time (s) & PPS (Gparticle/s) & CPP (cycles/particle) & Speedup ($\times$) \\
\midrule
\multicolumn{5}{l}{\textit{Interpolation \& Push}} \\
G0 & 131.312 & 1.635 & 0.795 & 1.00 \\
G1 & 42.476  & 5.056 & 0.257 & 3.09 \\
G2 & 27.778  & 7.731 & 0.168 & 4.73 \\
G3 & 22.936  & 9.363 & 0.139 & 5.73 \\
G4 & 22.013  & 9.756 & 0.133 & 5.97 \\
G5 & 25.121  & 8.548 & 0.152 & 5.23 \\
G6 & 21.178  & 10.140 & 0.128 & 6.20 \\
\textbf{G7} & \textbf{16.490} & \textbf{13.023} & \textbf{0.100} & \textbf{7.96} \\
\midrule
\multicolumn{5}{l}{\textit{Deposition}} \\
D0 & 86.221 & 2.491  & 0.522 & 1.00 \\
D1 & 51.142 & 4.199  & 0.310 & 1.69 \\
D2 & 6.922  & 31.023 & 0.042 & 12.46 \\
\textbf{D3} & \textbf{6.523} & \textbf{32.922} & \textbf{0.039} & \textbf{13.22} \\
\bottomrule
\end{tabular}%
}
\end{table}

Figure~\ref{fig:ablation_a_sve}(b) reveals that in the ultra-sparse regime ($PPC=1$), MPU variants lag due to unamortized padding and metadata overheads ($>80\%$ of runtime). However, as density increases ($PPC \ge 8$), the outer-product operator delivers substantial gains. G5 improves the kernel-stage throughput by $18.5\times$ over G0 at $PPC=512$, and introducing physical reordering (G6) further boosts memory bandwidth utilization by $1.2\times$. The fully realized G7, which combines \textit{Tail Sorting} with the \textit{Stream-Split} mechanism, achieves the optimal synergy between data layout and hardware characteristics. As detailed in Table~\ref{tab:ablation_combined_summary}, G7 reduces the computational cost to a minimal 0.100 CPP, representing a $7.96\times$ efficiency gain over the Baseline. These results confirm that while the MPU reformulation represents a significant architectural enhancement, the SoW-based data supply is the necessary enabler for its efficiency.

\begin{figure}[h]
\centering
\includegraphics[width=\linewidth]{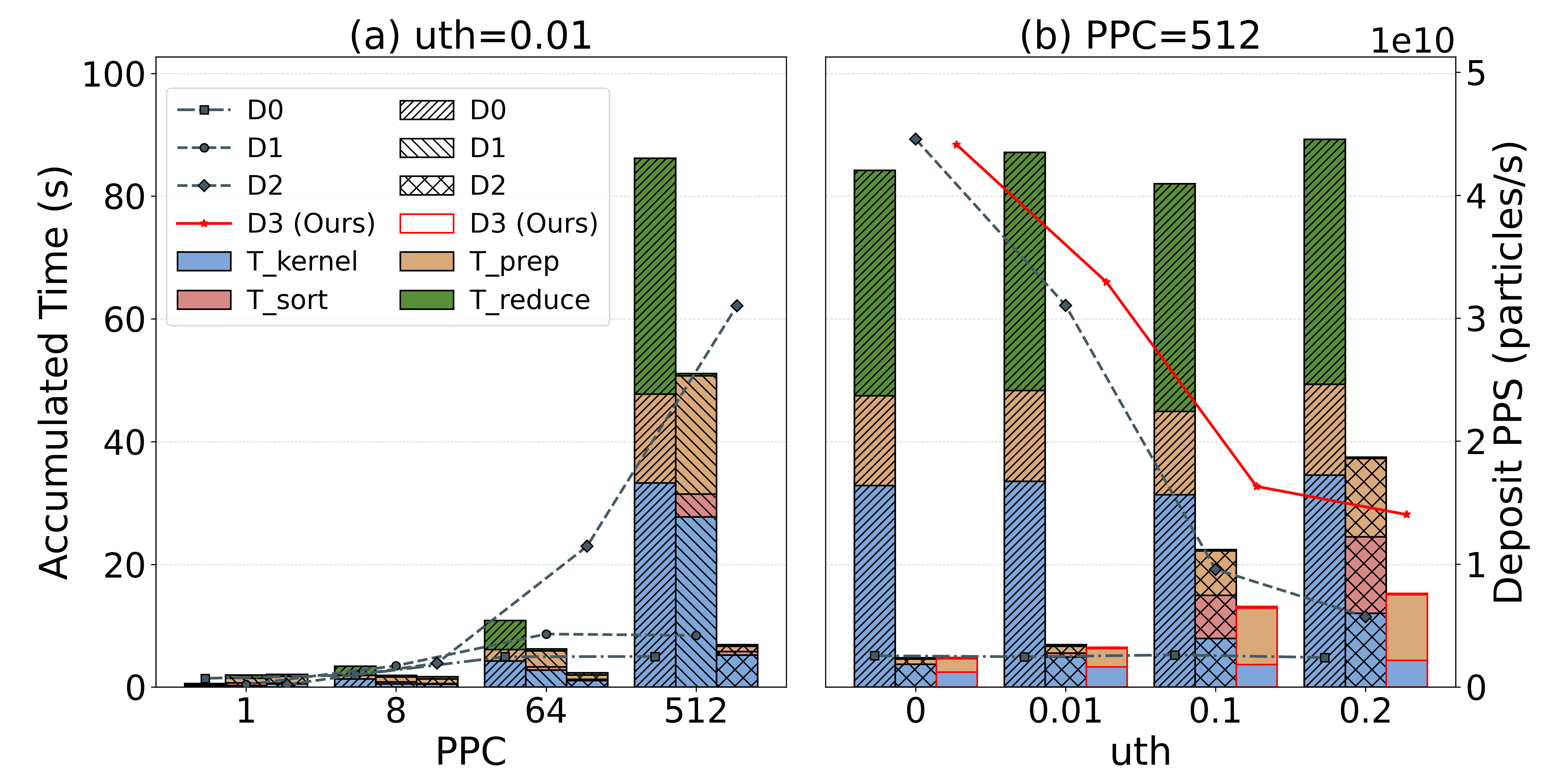}
\caption{Reuse benefits for D0--D2 under fixed $u_{th}=0.01$ and Robustness for D0, D2-D3 under PPC=512.}
\label{fig:ablation_b_reuse}
\end{figure}

\begin{figure*}[t]
    \centering
    \includegraphics[width=\textwidth]{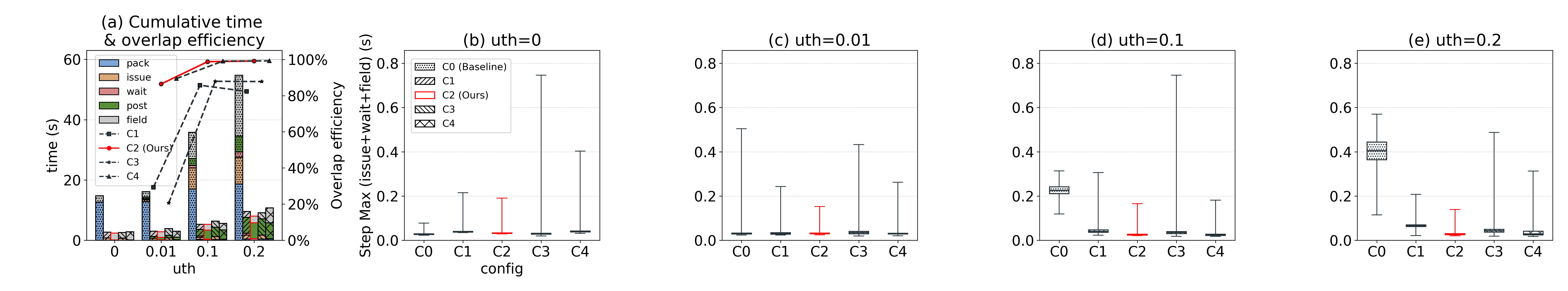}
    \caption{Impact of $u_{th}$ on overlap efficiency and runtime breakdown. Comparisons between average and maximum rank times highlight load imbalance and tail latency effects.}
    \label{fig:ablation_c_overlap}
\end{figure*}

\subsection{Ablation 2: Synergistic Gains in Deposition}

These experiments fix the Interpolation operator to either G2 (logical) or G4 (physical) and vary Deposition implementations (D0--D3) to evaluate the efficiency of layout reuse.

\subsubsection{Comparison of Different Layout Reuse (D0--D2)}

We evaluate layout reuse efficiency, finding that explicitly reusing the physically contiguous layout (D2) significantly outperforms logical ordering (D1), accelerating preparation and kernel execution by $23\times$ and $5.3\times$, respectively, at $PPC=512$.

Figure~\ref{fig:ablation_b_reuse}(a) shows that while both D1 and D2 mitigate atomic write conflicts compared to the baseline D0, the advantage of physical contiguity in D2 is decisive. Crucially, since D2 reuses the dual-region SoW structure and requires only lightweight sorting of the disordered tail, its sorting cost is reduced by $83.5\%$ relative to D1. This architectural efficiency is quantified in Table~\ref{tab:ablation_combined_summary}: while logically ordered variants consume $>0.3$ CPP due to atomic contention and index maintenance, the physically contiguous D2 effectively saturates Deposition bandwidth, reducing the cost to $\sim0.04$ CPP.

\subsubsection{Trade-off Between Dynamic--Static Splitting and Tail Over-Sorting (D0, D2--D3)}

Comparing tail handling strategies under varying migration intensities reveals that the dynamic--static splitting policy (D3) yields the optimal trade-off, avoiding the $33\%$ sorting overhead incurred by forced reordering (D2) at high turbulence.

We fix $PPC=512$ and sweep $u_{th}$ to evaluate robustness. As shown in Figure~\ref{fig:ablation_b_reuse}(b), D2 and D3 are comparable in low-migration regimes. However, as migration intensity increases, D2 suffers from escalating sorting costs and degraded MPU data supply efficiency for the resorted tail. At $u_{th}=0.2$, D2's kernel performance lags behind D3 by $2.7\times$. By falling back to the VPU atomic path only for the sparse disordered tail, D3 avoids expensive re-sorting while maintaining high-throughput MPU execution for the majority of particles, which reside in the Ordered Region. This confirms that reusing the physically contiguous layout combined with a hybrid splitting policy constitutes the robust Deposition pipeline even under worst-case migration scenarios.

\subsection{Ablation 3: Benefits of Overlap}

\subsubsection{Overlap Efficiency and Hardware Offloading (C0--C2)}

Evaluating communication efficiency under dynamic loads ($PPC=512$, varying $u_{th}$) reveals that the synergistic combination of SoW-based pre-packing and RDMA offloading (C2) eliminates explicit synchronization barriers, achieves a peak overlap ratio of $99.1\%$, and reduces total communication time by $20\times$ compared to the MPI-based alternative.

As illustrated in Figure~\ref{fig:ablation_c_overlap}(a), the foundation of this efficiency is the SoW mechanism, which eliminates redundant traversals and reduces the exposed standalone scan-and-pack stage by $\ge 99\%$ across all migration rates. Importantly, this does not mean that packing work vanishes entirely. Instead, the residual packing cost is absorbed into the write-back path described in Section~\ref{sec:repack} and is therefore accounted for within $T_{Interpolation}$ rather than as a separate end-of-step stage. As indicated by the measurements summarized in Figure~\ref{fig:ablation_a_sve}(a), the packing cost $T_{Sort}$ absorbed into the write-back store path contributes only 0.146\,s (3.3\% of $T_{Interpolation}$) to $T_{Interpolation}$ under the representative case of $u_{th}=0.01$, whereas the exposed standalone packing stage in the baseline accounts for 12.869\,s. Building on this fusion, the computation--communication pipeline further amplifies the gains. While the MPI-based C1 reaches a saturation point of 82.3\% overlap due to software overheads, the UNR-based C2 leverages RDMA hardware offloading to fully decouple data transmission from CPU intervention. This near-perfect overlap effectively masks redistribution costs. Specifically, C2 reduces the maximum per-rank wait time to just 0.01\,s, representing a $58\times$ reduction relative to C1, which verifies that the asynchronous design successfully hides synchronization latency behind the Deposition kernel.

\subsubsection{Stability and Network Contention (C0--C4)}

Analyzing the stability of the communication-critical path ($T_{issue} + T_{wait} + T_{field}$) via boxplots reveals that the conservative overlap strategy (C2) provides the most robust time-to-solution by avoiding the network contention inherent in aggressive  overlap strategies (C3/C4).

As shown in Figure~\ref{fig:ablation_c_overlap}(b--e), while the bulk-synchronous baseline (C0) suffers a scalability collapse with a $>14\times$ latency surge under dynamic loads ($u_{th}=0.2$), asynchronous strategies generally mitigate this impact. However, attempting to maximize overlap by extending particle traffic into the field-solver phase (C3/C4) proves counterproductive. Specifically, although the MPI-based C3 lowers median latency, it induces severe jitter, increasing maximum latency by $2.3\times$ compared to C1. This contention persists even with hardware offloading: the aggressive C4 exhibits a $2.25\times$ longer tail compared to the conservative C2, despite having comparable medians. This degradation confirms that extending particle traffic saturates NIC bandwidth, interfering with latency-sensitive field synchronization. By isolating communication to the Deposition window, our method (C2) eliminates these resource conflicts, maintaining the tightest latency distribution and ensuring predictable time-to-solution.

\subsection{Cross-Platform Peak Efficiency}

For contextual reference, POLAR-PIC achieves $13.2\%$ of theoretical peak efficiency on the LX2 CPU, while WarpX reaches $9.6\%$ on the NVIDIA A800 GPU. On the evaluated LX2 platform, this corresponds to a $13.2\times$ improvement in peak-efficiency ratio over the native WarpX port.

As detailed in Table~\ref{tab:peak_efficiency}, the native WarpX reference on LX2 is constrained to $\sim1\%$ efficiency due to irregular memory access and VPU throughput ceilings, whereas POLAR-PIC substantially improves utilization through outer-product reformulation and sustained data continuity. On the evaluated matrix-centric CPU platform, this domain-specific co-optimization yields a $2.4\times$ gain over Matrix-PIC and achieves a node-level Figure of Merit ($\text{FOM}_{\text{node}}$) of $1.1\times10^{10}$. We report the cross-platform comparison only as context, since the platforms and software stacks differ.

\begin{table}[h]
  \centering
  \begin{threeparttable}
    \caption{Cross-platform performance comparison (double precision; normalized to platform peak and reported for context).}
    \label{tab:peak_efficiency}
    \footnotesize
    \setlength{\tabcolsep}{4pt}
    \begin{tabular}{l l c c}
      \toprule
      Platform & Variant & $\eta_{\mathrm{peak}}$ (\%) & $\mathrm{FOM}_{\mathrm{node}}$ \\
      \midrule
      \multirow{3}{*}{LX2 CPU (This work)\tnote{*}}
        & WarpX (Native)          & \phantom{0}1.0  & 8.3e8   \\
        & Matrix-PIC              & \phantom{0}5.5  & 3.6e9   \\
        & \textbf{POLAR-PIC}      & \textbf{13.2} & \textbf{1.1e10} \\
      \midrule
      NVIDIA A800 (Measured)\tnote{*}
        & WarpX (CUDA)            & \phantom{0}9.6  & 3.3e9   \\
      \midrule
      \multirow{4}{*}{Reference Platforms\tnote{**}}
        & WarpX (Perlmutter A100) & 12.9 & 9.2e8   \\
        & WarpX (Summit V100)     & \phantom{0}8.3  & 8.0e8   \\
        & WarpX (Frontier MI250X) & \phantom{0}3.3  & 1.3e9 \\
        & WarpX (Fugaku A64FX)    & \phantom{0}1.1  & 2.2e7   \\
      \bottomrule
    \end{tabular}
    \begin{tablenotes}[flushleft]
      \scriptsize
      \item[*] $\eta_{\mathrm{peak}}$ is normalized to the theoretical peak performance of the target hardware.
      \item[**] Reference values from~\cite{fedeli2022warpx} are provided for context.
    \end{tablenotes}
  \end{threeparttable}
\end{table}

\subsection{Weak Scalability at Scale}
\label{sec-exp:scale}

In the end-to-end full-timestep weak-scaling study from 1 to 4,096 nodes (over 2 million physical cores) under high-turbulence stress ($u_{th}=0.2$), POLAR-PIC maintains a robust weak-scaling efficiency of $67.5\%$, significantly outperforming the native WarpX reference, which drops to $42.5\%$ due to unmasked communication overheads.

\begin{figure}[t]
\centering
\includegraphics[width=\linewidth]{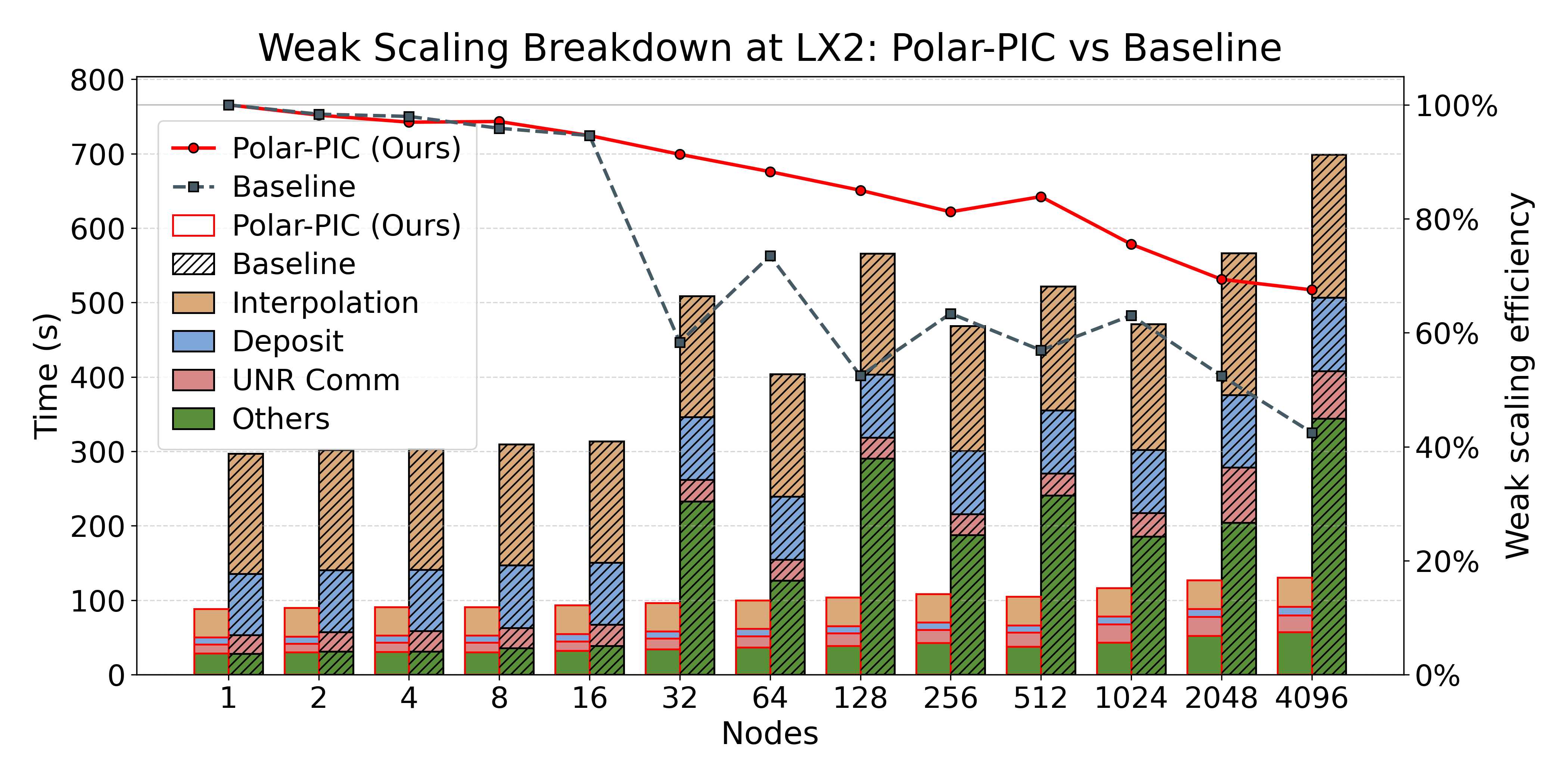}
\caption{End-to-end full-timestep weak-scaling breakdown from 1 to 4096 nodes (over 2 million cores) on the LS system under high-turbulence stress ($PPC=512,\ u_{th}=0.2$).}
\label{fig:weak_scaling}
\end{figure}

Figure~\ref{fig:weak_scaling} reveals the divergence in end-to-end scaling behavior. While both implementations exhibit scalability exceeding $90\%$ up to 16 nodes, the native WarpX reference degrades sharply at scale because blocking communication fails to hide the growing latency of large-scale interconnects and heavy particle redistribution overheads. In contrast, POLAR-PIC’s asynchronous pipeline effectively masks these overheads, sustaining near-ideal scaling ($\sim100\%$) for the particle-handling components (Interpolation, Deposition, and Particle Redistribution) even at the 2-million-core scale, whereas the native WarpX particle efficiency drops to $62.8\%$. These results validate that our co-design substantially alleviates the particle-processing bottleneck. However, the end-to-end breakdown also reveals an Amdahl's Law implication: once particle costs are reduced, the Field Solver becomes the dominant residual cost, identifying the field-update phase as the critical target for future optimization.

\section{Conclusion and Future Work}

This paper presents \textbf{POLAR-PIC}, a holistic co-design framework that systematically resolves the architectural mismatches between traditional PIC algorithms and emerging matrix-centric hardware. By synergizing the reformulation of Field Interpolation into an MPU-friendly outer-product kernel, the enforcement of physical memory contiguity via the Sort-on-Write (SoW) mechanism, and the deployment of an RMA-based asynchronous communication pipeline, POLAR-PIC establishes a highly efficient integrated execution path for particle simulations.

Experimental evaluations on the next-generation LS pilot system demonstrate that POLAR-PIC significantly improves the particle-processing path over existing solutions. In uniform plasma benchmarks, it achieves speedups of up to $10.9 \times$ over the native WarpX reference pipeline on LX2 and outperforms the matrix-based state-of-the-art Matrix-PIC by $4.7 \times$. Extending to real-world laser-ion acceleration scenarios, the framework maintains robust performance, achieving speedups of $4.4\times$ over the native WarpX reference and $3.8\times$ over Matrix-PIC despite intense particle migration. Moreover, the asynchronous communication design effectively masks redistribution overhead, sustaining an overlap ratio of $99.1\%$ even under high-turbulence conditions. For contextual reference, POLAR-PIC attains $13.2\%$ of theoretical peak efficiency on the evaluated CPU-based LS system, while WarpX reaches $9.6\%$ on NVIDIA A800 GPUs. Furthermore, the framework maintains $67.5\%$ end-to-end weak scaling efficiency from 1 to 4,096 nodes (aggregating over 2 million cores).

With the particle-processing bottleneck effectively mitigated, our end-to-end weak-scaling analysis indicates that the field solver emerges as the new dominant cost in the simulation time step. Consequently, future work will focus on optimizing field-solver algorithms and their associated communication paths to better match the throughput of the accelerated particle kernels. Additionally, we plan to extend the POLAR-PIC paradigm to support more complex multi-physics scenarios, such as Quantum Electrodynamics (QED) and dynamic load balancing in non-uniform mesh refinements, to address a broader spectrum of high-energy physics challenges.

\begin{acks}
We sincerely thank all the anonymous reviewers for their valuable feedback. This research was supported by Guangdong S\&T Program under Grant No. 2024B0101040005, the National Natural Science Foundation of China (NSFC): No.62461146204 and No.62502552, Guangdong Province Special Support Program for Cultivating High-Level Talents: 2021TQ06X160, National Key Research and Development Program of China under Grant No. YFE030170000 and the Strategic Priority Research Program of Chinese Academy of Sciences under Grant No. XDB0790203.

This research used the open-source particle-in-cell code \href{https://github.com/BLAST-WarpX/warpx}{WarpX}. Primary WarpX contributors are with LBNL, LLNL, CEA-LIDYL, SLAC, DESY, CERN, Helion Energy, and TAE Technologies. We acknowledge all WarpX contributors.
\end{acks}

\bibliographystyle{unsrt}
\bibliography{HPDC-sigconf}

\appendix
\section{Implementation Details}
\label{append:implementation}
Table~\ref{tab:exp_config_impl} details the specific implementation strategies for each experimental variant (G0--G7, D0--D3, and C0--C4). It highlights the architectural distinctions between baseline VPU approaches and the proposed MPU-accelerated designs, including the specific sorting mechanisms and overlap protocols employed to isolate performance contributions.
\begin{table*}[t]
\centering
\caption{Detailed implementations for experimental configurations and ablation variants.}
\label{tab:exp_config_impl}
\small
\setlength{\tabcolsep}{6pt}
\renewcommand{\arraystretch}{1.2}
\begin{tabular}{l p{0.50\textwidth} p{0.38\textwidth}}
\toprule
\textbf{Config} & \textbf{Implementation Specification} & \textbf{Evaluation Purpose \& Significance} \\
\midrule
\multicolumn{3}{l}{\textit{\textbf{Exp 1: Interpolation (Gather \& Push) Variants}}} \\
G0 &
Native WarpX Gather; relies on compiler auto-vectorization for VPU; no particle ordering. &
Establishes the performance baseline under irregular memory access patterns. \\
G1 &
Hand-tuned VPU Gather using explicit SIMD intrinsics; unsorted. &
Isolates the benefits of manual VPU instruction optimization from data locality factors. \\
G2 &
G1 with incremental logical index sorting. &
Reproduces the logical-order supply strategy of Matrix-PIC~\cite{rao2026matrix}; quantifies locality benefits without physical data movement. \\
G3 &
G2 enhanced with physical reordering based on the sorted indices (explicit memory compaction). &
Decouples the impact of physical compaction from logical indexing under the VPU kernel. \\
G4 &
Replaces G3's reordering with Sort-on-Write (SoW) mechanisms during the write-back path. &
Evaluates the efficiency of SoW compared to explicit memory compaction. \\
\cmidrule(lr){1-3}
G5 &
MPU Gather kernel enabled; retains G2-style logical index sorting. &
Assesses MPU sensitivity to logical-order supply and memory fragmentation (MPU counterpart to G2). \\
G6 &
MPU Gather kernel with G3-style physical reordering driven by indices. &
Measures the synergistic effect of MPU execution and explicit physical compaction (MPU counterpart to G3). \\
G7 &
MPU Gather kernel with SoW physical reordering (Final POLAR-PIC Gather). &
Validates that SoW-sustained contiguity enables stable, sustained MPU pipeline saturation. \\
\midrule
\multicolumn{3}{l}{\textit{\textbf{Exp 2: Deposition (Scatter) Variants}}} \\
D0 &
Native WarpX Deposition; uses default atomic conflict handling strategies. &
Baseline for Deposition computational cost and conflict resolution without MPU reformulation. \\
D1 &
MPU Deposition reusing G2's incremental logical indices. &
Reproduces the Matrix-PIC Deposition strategy~\cite{rao2026matrix}; isolates MPU gains under logical ordering. \\
D2 &
MPU Deposition reusing G7's physical layout; adds extra binning for the disordered tail stream. &
Evaluates whether explicit tail binning improves robustness when perfect global contiguity is absent. \\
D3 &
MPU Deposition reusing G7's physical layout; falls back to VPU for the tail stream. &
Final design of POLAR-PIC. Maximizes reuse of SoW layout while maintaining efficient, low-overhead tail handling. \\
\midrule
\multicolumn{3}{l}{\textit{\textbf{Exp 3: Overlap Strategy Variants}}} \\
C0 &
Bulk-synchronous redistribution at end of the step (Scan $\to$ Pack $\to$ Send $\to$ Wait $\to$ Unpack). &
Baseline BSP behavior; exposes full packing and synchronization latency on the critical path. \\
C1 &
MPI-based nonblocking overlap; initiates transfers and converges immediately after Deposit. &
Evaluates conservative overlap capabilities using standard MPI progress semantics. \\
C2 &
UNR-based notifiable overlap; offloads pre-packed buffers and converges after Deposit. &
Default POLAR-PIC strategy. Validates NIC offloading and notifiable completion for stable overlap. \\
C3 &
MPI-based overlap extended into the Field Solve phase. &
Assesses aggressive overlap limits and potential bandwidth interference with field communication. \\
C4 &
UNR-based overlap extended into the Field Solve phase. &
Evaluates whether UNR enables deeper overlap windows without destabilizing overall performance. \\
\midrule
\multicolumn{3}{l}{\textit{\textbf{Integrated Systems}}} \\
Baseline &
Reference pipeline: (G0 + D0 + C0). &
Represents the production-grade WarpX configuration tuned with official optimizations for best-effort performance. \\
Matrix-PIC &
Open-source-based Matrix-PIC architecture (G2 + D1 + C0) reconstructed by reusing incremental index sorting logic.  & 
Extends the matrixized SOTA by applying logical index-sorting to both Interpolation and Deposition without SoW physical reordering. Represents the manually implemented best-case baseline SOTA without SoW. \\
POLAR-PIC &
Proposed pipeline: (G7 + D3 + C2). &
Integrates full MPU execution, SoW-sustained contiguity, and UNR-based overlap. \\
\bottomrule
\end{tabular}
\end{table*}

\section{Simulation Parameters}
\label{append:parameter}
Table~\ref{tab:workload_parameters_used} lists the comprehensive physical and numerical parameters used for the Uniform Plasma and Laser-Ion Acceleration benchmarks. These parameters are chosen to ensure the reproducibility of the performance results and to reflect representative workloads in high-energy density physics simulations.
\begin{table*}[t]
\centering
\caption{Key parameters for Uniform Plasma and Laser-Ion Acceleration workloads.}
\label{tab:workload_parameters_used}
\small
\setlength{\tabcolsep}{5pt}
\renewcommand{\arraystretch}{1.2}
\begin{tabularx}{\textwidth}{l >{\raggedright\arraybackslash}X >{\raggedright\arraybackslash}X}
\toprule
\textbf{Input Parameter} & \textbf{Uniform Plasma} & \textbf{Laser-Ion Acceleration} \\
\midrule
\textbf{Objective} &
Assess SIMD/MPU kernel efficiency under controlled conditions &
Evaluate system-level behavior under strongly non-uniform, migration-heavy dynamics \\
\textbf{Physics Context} &
Homogeneous periodic plasma for isolating per-step kernel behavior &
Laser--solid interaction capturing transient acceleration dynamics \\
\midrule

\multicolumn{3}{l}{\textit{\textbf{Simulation Configuration}}} \\
\quad \texttt{max\_step} & 100 & 100 \\
\quad \texttt{geometry.dims} & 3 & 3 \\
\quad \texttt{warpx.grid\_type} & collocated & collocated \\
\quad \texttt{amr.n\_cell} & $256 \times 128 \times 128$ & $192 \times 192 \times 256$  \\
\quad \texttt{geometry.prob\_lo} & $-2.0 \times 10^{-5}$ (all dims) & $-3.75 \times 10^{-6}, -3.75 \times 10^{-6}, -2.5 \times 10^{-6}$ \\
\quad \texttt{geometry.prob\_hi} & $+2.0 \times 10^{-5}$ (all dims) & $+3.75 \times 10^{-6}, +3.75 \times 10^{-6}, +1.25 \times 10^{-5}$ \\
\midrule

\multicolumn{3}{l}{\textit{\textbf{Boundary Conditions \& Solvers}}} \\
\quad \texttt{boundary.field} & \texttt{periodic} (all faces) & \texttt{pml} (all faces) \\
\quad \texttt{warpx.cfl} & 1.0 & 0.999 \\
\quad \texttt{algo.particle\_shape} & 3 (Cubic) & 3 (Cubic) \\
\quad \texttt{algo.maxwell\_solver} & Yee & Yee \\
\quad \texttt{algo.particle\_pusher} & Boris & Boris \\
\midrule

\multicolumn{3}{l}{\textit{\textbf{Particle Configuration}}} \\
\quad \texttt{particles.tile\_size} & $8 \times 8 \times 8$ & $8 \times 8 \times 8$ \\
\quad \texttt{electrons.profile} & Constant & \texttt{parse\_density\_function} \\
\quad \texttt{electrons.density} & $1.0 \times 10^{25} \ \mathrm{m}^{-3}$ & -- \\
\quad \texttt{electrons.ppc} & \{1, 2, 4, \dots, 512\} & \{1, 2, 4, \dots, 512\} \\
\quad \texttt{electrons.u\_th} (x,y,z) & $\{0, 0.01, 0.05, 0.1, 0.2\}$ & $0.01$ (fixed) \\
\quad \texttt{target parameters} & -- & $L=50\,\mathrm{nm}, n_e=30n_c$ \\
 & & $n_c=1.74 \times 10^{27} \ \mathrm{m}^{-3}$ \\
\midrule

\multicolumn{3}{l}{\textit{\textbf{Laser Parameters (LIA only)}}} \\
\quad \texttt{laser.profile} & -- & Gaussian \\
\quad \texttt{laser.a0} & -- & 16.0 \\
\quad \texttt{laser.wavelength} & -- & $0.8 \ \mu\mathrm{m}$ \\
\quad \texttt{laser.duration} & -- & $30 \ \mathrm{fs}$ \\
\quad \texttt{laser.focal\_dist} & -- & $4.0 \ \mu\mathrm{m}$ \\
\quad \texttt{laser.waist} & -- & $4.0 \ \mu\mathrm{m}$ \\
\bottomrule
\end{tabularx}
\end{table*}

\section{Artifact Appendix}

\subsection{Abstract}
This artifact provides the open-source implementation of \textbf{POLAR-PIC}
The released repository is built on top of WarpX and contains the implementation of the main ideas presented in this paper: an MPU-oriented particle-processing path, physically ordered particle layout management, and overlapped particle redistribution.

\subsection{Description \& Requirements}

\subsubsection{How to access}
The artifact is publicly available at
\url{https://github.com/sherry-roar/polarpic-ad}.

\subsubsection{What is released}
The repository releases the implementation basis of POLAR-PIC on top of WarpX. In particular, it includes the code paths corresponding to:
\begin{itemize}
    \item matrixized / physically ordered particle push,
    \item custom order-3 deposition,
    \item communication overlap and one-sided communication integration,
    \item AMReX-side metadata changes required by physical sorting and incremental sorting.
\end{itemize}
At the algorithm level, the released implementation reflects the three key ideas of the paper: reformulating Field Interpolation into an MPU-friendly outer-product form, maintaining physical locality through particle layout management, and overlapping particle communication with computation.

\subsubsection{Dependencies}
The artifact is based on WarpX and retains the corresponding WarpX/AMReX 24.07 implementation context. It also includes a vendored one-sided communication backend used by the released implementation. The best performance reported in the paper is obtained on the LS pilot system with LX2 CPUs and the associated software stack described in the main paper.

\subsubsection{Portability}

The repository preserves the source-level implementation of the POLAR-PIC optimization path, while some low-level mappings remain platform dependent. In the released code, the \texttt{vpu}/\texttt{mpu} paths should be understood as placeholders for the target machine's vector/matrix execution path. When porting to another architecture, these parts can be adapted by replacing the corresponding \texttt{vpu}/\texttt{mpu} instruction functions and related backend-specific support. For example, when targeting Apple M-series processors, the same optimization path can in principle be retargeted by replacing the current \texttt{vpu}/\texttt{mpu} implementations with the corresponding instruction-level functions on that platform. Likewise, if the original communication backend is unavailable, the communication path can be reimplemented with an equivalent one-sided interface, for example using MPI one-sided communication semantics, provided that the required synchronization behavior is preserved.

\subsubsection{Additional note}
The released artifact is intended to expose the implementation of POLAR-PIC in an inspectable open-source form. The strongest performance results in this paper correspond to the LS/LX2 platform. If evaluators or readers have questions about the code or platform adaptation, they may contact the authors.
\end{document}